\documentclass[11pt]{article}

\usepackage{acl}

\usepackage{times}
\usepackage{latexsym}

\usepackage[T1]{fontenc}

\usepackage[utf8]{inputenc}

\usepackage{microtype}

\usepackage{inconsolata}

\usepackage{graphicx}

\usepackage{booktabs}
\usepackage{amsmath}
\usepackage{amssymb}
\usepackage{bm} 
\usepackage{esint} 
\usepackage{commath} 
\usepackage{nicefrac} 
\usepackage{multirow} 
\usepackage{color}
\usepackage[table]{xcolor}

\renewcommand{\mid}{\,\ifnum\currentgrouptype=16 \middle\fi|\,}

\usepackage{pifont}
\newcommand{\cmark}{\ding{51}}%
\newcommand{\xmark}{\ding{55}}%

\title{Affectron: Emotional Speech Synthesis with Affective and Contextually Aligned Nonverbal Vocalizations}

\author{
  Deok-Hyeon Cho, Hyung-Seok Oh, Seung-Bin Kim, Seong-Whan Lee\thanks{Corresponding author} \\
  Department of Artificial Intelligence, Korea University, Seoul, Korea\\
  \texttt{\{dh\_cho,hs\_oh,sb-kim,sw.lee\}@korea.ac.kr}}

\begin{document}
\maketitle
\begin{abstract}
Nonverbal vocalizations (NVs), such as laughter and sighs, are central to the expression of affective cues in emotional speech synthesis. However, learning diverse and contextually aligned NVs remains challenging in open settings due to limited NV data and the lack of explicit supervision. Motivated by this challenge, we propose Affectron as a framework for affective and contextually aligned NV generation. Built on a small-scale open and decoupled corpus, Affectron introduces an NV-augmented training strategy that expands the distribution of NV types and insertion locations. We further incorporate NV structural masking into a speech backbone pre-trained on purely verbal speech to enable diverse and natural NV synthesis. Experimental results demonstrate that Affectron produces more expressive and diverse NVs than baseline systems while preserving the naturalness of the verbal speech stream.
\end{abstract}

\section{Introduction}
Nonverbal vocalizations (NVs), including laughter, sighs, and cries, are essential for conveying affect in human communication, complementing prosodic modulation \cite{cortes2021effects, kanda2024making, wu2024laugh, ye2025scalable}. However, most expressive text-to-speech (TTS) systems \cite{lee2022hierspeech, im2022emoq, zhou2022emotion, zhou2023speech, inoue2024hierarchical, cho25b_interspeech, wang2025word, gao2025emo, gudmalwar2025emoreg} remain limited in their ability to generate expressive speech that incorporates natural NVs.

Existing studies on speech synthesis incorporating NVs can be broadly categorized into two main approaches. The first is tag-controlled TTS \cite{zhang2023nsv, kanda2024making, wu2024laugh, borisov2025nonverbaltts, ye2025scalable}, which manually inserts explicit tags (e.g., $\left< \text{laughing} \right>$, $\left< \text{filler} \right>$) to specify NV type and location. Although this approach enables fine-grained control, it relies on aligned annotations or NV detection models \cite{omine24_interspeech, schmid2025effective}, whose biases and error propagation often result in temporal inconsistencies in NV locations. The second is spontaneous-style TTS \cite{li2024spontts, li24na_interspeech}, which predicts NVs from contextual cues without explicit alignment. Reproducibility in this approach is constrained by reliance on proprietary datasets and by limitations in the scale and quality of publicly available corpora. Several NV-integrated datasets have been introduced in recent work to mitigate these issues. However, richly annotated corpora \cite{zhang2023nsv, li2024spontts} are generally not publicly available. In contrast, publicly available corpora \cite{xin2024jnv, xin2024jvnv, wang2025capspeech, liao2025nvspeech, borisov2025nonverbaltts} are generally skewed toward basic NVs, such as breathing and laughter, and frequently exhibit acoustic artifacts. Consequently, the absence of large, diverse, and high-quality public NV corpora continues to hinder accurate modeling of fine-grained NV variations, such as subtle chuckles, giggles, and snickers.

Recent developments in generative speech modeling have significantly enhanced the naturalness and flexibility of TTS systems. Neural codec language model (NCLM)-based zero-shot TTS systems \cite{peng2024voicecraft, wang2023neural, du2024cosyvoice, huang25c_interspeech} can be trained on diverse and even low-quality speech corpora while still producing highly natural synthesized speech. Nevertheless, current NCLM-based TTS systems primarily focus on voice cloning, and the generation of human-like expressive speech with integrated NVs remains insufficiently explored. As a result, the ability to control NVs with fine-grained prosodic variation remains limited.

To address these limitations, we introduce Affectron, a method for generating affectively and contextually aligned NVs. During training, we mitigate data scarcity and bias by leveraging a small-scale open and decoupled corpus in which verbal speech and NVs are recorded separately \cite{richter24_interspeech}. Based on this setting, we propose an NV-augmented training strategy that enhances the capability of a verbal-only pre-trained NCLM \cite{peng2024voicecraft} to model diverse NV types and insertion locations. This augmentation comprises two components: (i) we introduce emotion-driven top-$K$ NV matching, which selects emotionally aligned NVs for each verbal utterance to enhance affective consistency and increase NV diversity. (ii) we propose emotion-aware top-$K$ routing, which locates the selected NVs at contextually appropriate locations, thereby eliminating dependence on alignment annotations or NV detectors. These two modules are used only during training to construct NV-augmented samples. Additionally, we incorporate NV structural masking into the NCLM to condition generation on the affective context of the surrounding verbal speech. During inference, the model generates speech from an NV-tagged text and an emotional reference utterance, without requiring any matching or routing procedure. Experimental results demonstrate that Affectron produces more expressive and diverse NV synthesis than previous NCLM-based TTS systems, while maintaining the naturalness of the verbal stream. Moreover, the proposed augmentation yields NV type-location distributions that more closely align with empirical data compared to competing approaches. Our audio samples and implementation code are publicly available at \url{https://choddeok.github.io/Affectron/}. 

\section{Related Work}
\subsection{Speech Synthesis with NVs}
Recent studies have explored the generation of NVs in emotional TTS systems. Laughter Synthesis \cite{xin23b_interspeech} feeds pseudo-phonetic tokens into the TTS to produce laughter.  However, this method determines the location and variation of laughter in a largely stochastic manner, which limits controllability and constrains the expressive range. ELaTE \cite{kanda2024making} and EmoCtrl-TTS \cite{wu2024laugh} condition flow-matching-based TTS systems on NV embeddings. However, these works attempt to reduce labeling cost using NV detectors \cite{omine24_interspeech}, but these methods are generally applicable only to laughter and crying, which restricts their generalizability. Spontaneous-TTS \cite{li24na_interspeech} supplies behavior labels and syntactic cues derived from linguistic features to model NVs explicitly. However, this approach depends on proprietary labeled datasets and necessitates explicit supervised annotations for training. Meanwhile, CosyVoice \cite{du2024cosyvoice, du2025cosyvoice} successfully synthesizes natural speech with NVs and supports fine-grained control. These methods require extensive, high-quality annotated corpora, and the resulting synthesized speech often lacks naturalness when multiple NV types are generated concurrently.

\subsection{Neural Codec Language Models}
NCLMs have recently emerged as a prominent paradigm for speech generation. They discretize audio into codec tokens, model sequential dependencies via next-token prediction, and decode the predicted tokens back into high-quality audio. Recent zero-shot TTS approaches frequently employ NCLMs as the foundational architecture. A representative example is VALL-E \cite{wang2023neural}, which predicts part of the EnCodec codebooks \cite{defossez2022high} using an autoregressive (AR) codec language model, while generating the remaining codebooks with a non-AR model. This design demonstrates that zero-shot TTS can be achieved by conditioning a codec language model on a brief reference prompt. VoiceCraft \cite{peng2024voicecraft} is a Transformer-based NCLM that performs AR token infilling under a bidirectional context using a two-stage token rearrangement. This architecture facilitates efficient multi-codebook modeling and exhibits strong performance in both speech editing and zero-shot TTS. However, current NCLM-based TTS systems are primarily designed for voice cloning and have not been extensively investigated for generating human-like expressive speech. These systems frequently fail to reliably generate NVs with fine-grained prosodic variations, even when such cues are explicitly provided in the prompt.

\section{Background}
\subsection{Affective Dynamics for NV Type and Location}
NVs convey affective states more directly and effectively than verbal prosody, serving as subtle emotional cues in speech. Incorporating NVs into a speech synthesis system requires careful selection of NV types that align with the surrounding affective context \cite{gupta2012classification, banninger2023different}. In addition to selecting appropriate NV types, the location of NV occurrences significantly influences the perception and integration of emotional cues with verbal speech. Previous research \cite{ephratt2008functions, hoey2014sighing} demonstrates that the location of NVs within a sentence substantially affects the expressiveness and immersive quality of emotional communication.

\begin{figure}[!t]
  \centering
\centerline{\includegraphics[width=0.4\textwidth]{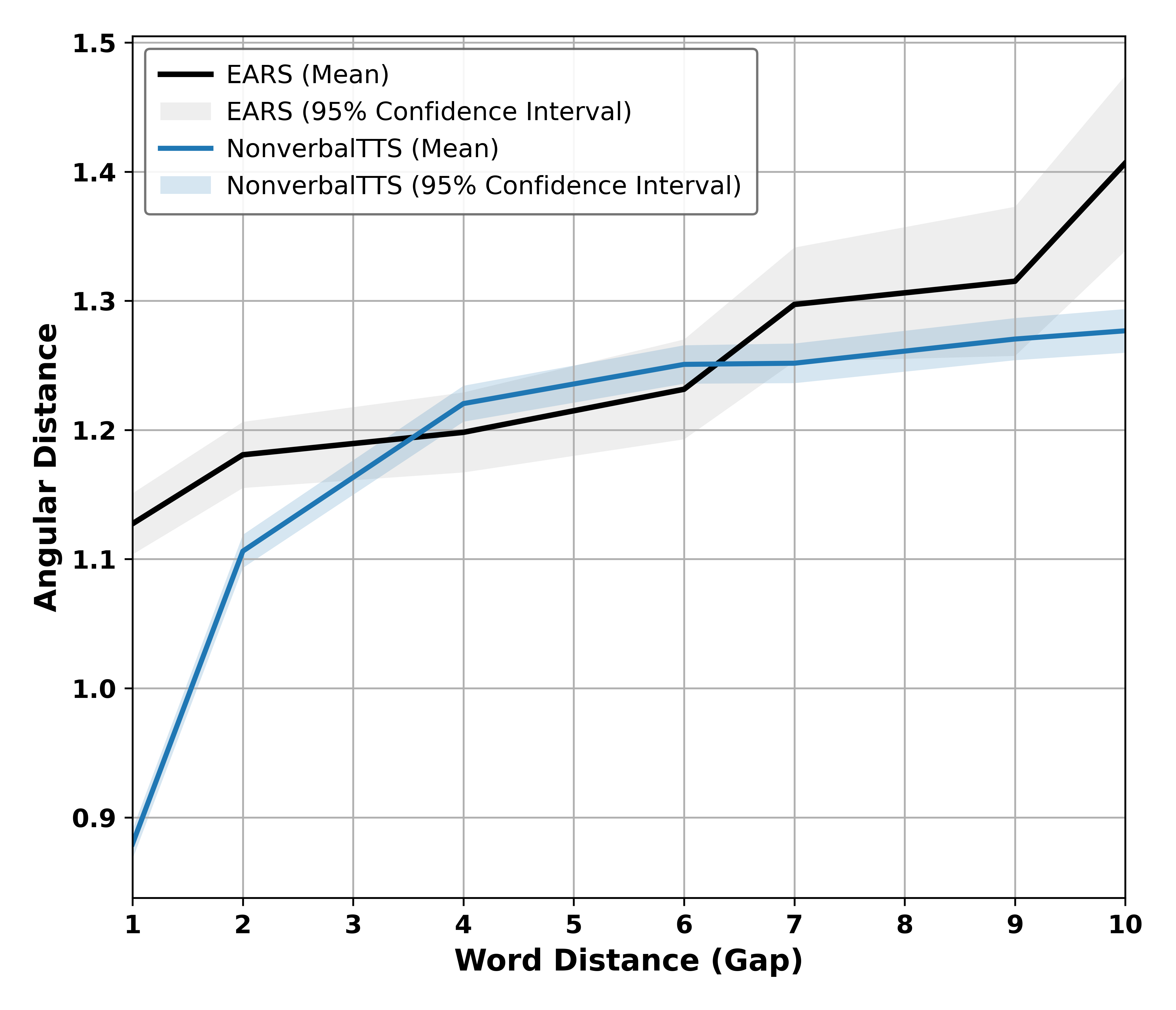}}\vspace{-0.2 cm}
\caption{Analysis of emotional attribute change patterns across temporal gaps (up to a gap of 10) for the EARS \cite{richter24_interspeech} dataset (verbal only) and the NonverbalTTS \cite{borisov2025nonverbaltts} dataset (verbal-nonverbal combined speech).}
\label{fig:emotion_gap} \vspace{0.0cm}
\end{figure}

Furthermore, studies on continuous emotion modeling \cite{wollmer2013lstm, huang2018prediction, liu2025emo} indicate that emotional states typically evolve gradually rather than shift abruptly over time. Motivated by these characteristics of emotional transitions, we analyzed temporal patterns of emotional attribute changes across consecutive word segments in real data \cite{richter24_interspeech, borisov2025nonverbaltts}. To compute these patterns, emotional attribute pseudo-labels \cite{wagner2023dawn} were transformed into spherical coordinates \cite{cho24_interspeech, 10965917, cho25_interspeech}, and angular distances were measured on the unit sphere. This representation highlights directional changes in affective dynamics, enabling angular distance to more reliably capture relative emotional transitions between adjacent segments. As shown in Figure \ref{fig:emotion_gap}, shorter temporal intervals between words are associated with smaller angular distances. These findings indicate that affective states change gradually and exhibit local stability over brief temporal spans. Based on this observation, locations exhibiting minimal emotional attribute change are identified as emotionally stable locations and are assumed to serve as natural anchor points for NV event insertion. In our framework, inserting NVs at such locations is expected to preserve affective coherence while enhancing expressiveness. Appendix \ref{appendix:emotion_dynamics} provides a more detailed analysis of these temporal patterns of emotional attributes.

\subsection{VoiceCraft Overview}
\label{sec:voiceCraft}
VoiceCraft \cite{peng2024voicecraft} serves as the backbone of Affectron, leveraging causal masking and delayed stacking to enable context-aware editing and infilling. Causal masking \cite{aghajanyan2022cm3, donahue-etal-2020-enabling, bavarian2022efficient} moves a masked span to the end of the sequence, allowing the model to condition on both past and future context. This bidirectional conditioning improves boundary naturalness and contextual coherence. Based on this, we hypothesize that bidirectional conditioning can benefit NV synthesis by leveraging the surrounding affective context. Delayed stacking \cite{copet2023simple} enables efficient multi-codebook AR modeling by adding a cumulative delay across EnCodec codebook streams \cite{defossez2022high}. This mechanism facilitates high-fidelity token prediction across parallel acoustic channels and supports the generation of fine-grained, high-quality audio signals. Therefore, high-fidelity token modeling is expected to be particularly advantageous for NV synthesis, which poses greater modeling challenges than verbal speech.

A decoder-only Transformer is trained to predict speech tokens, including the masked spans, conditioned on a transcript and optimized with a cross-entropy loss \cite{aghajanyan2022cm3}. Training on both observed and masked regions provides supervision at every timestep, stabilizing optimization and accelerating convergence.

\begin{figure*}[!t]
  \centering
\centerline{\includegraphics[width=1.0\textwidth]{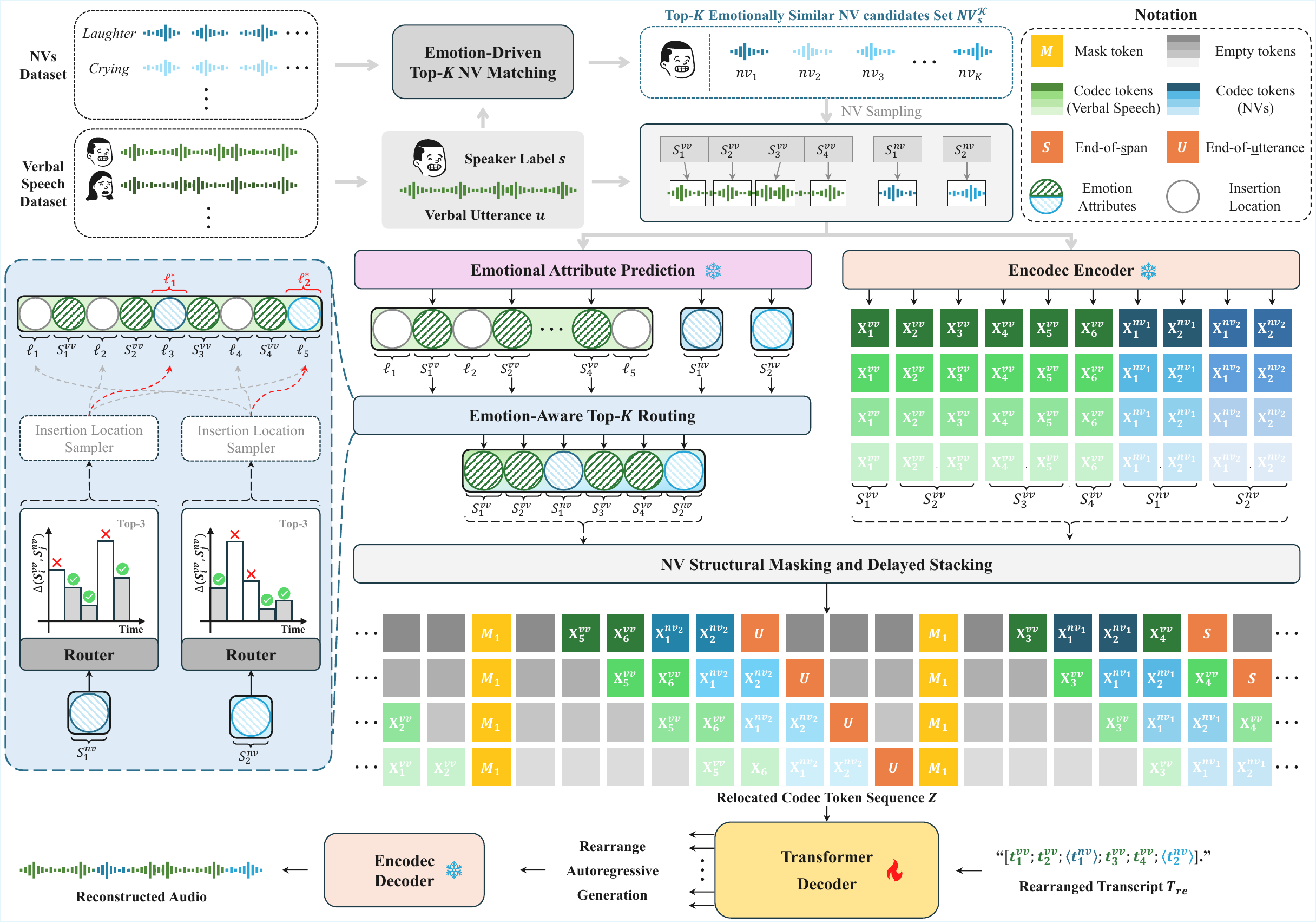}}\vspace{-0.2 cm}
\caption{Overview of the Affectron training framework. NV candidates are selected and routed to contextually appropriate locations to construct NV-augmented training samples, which are then used to fine-tune the VoiceCraft backbone for affect-aware NV synthesis.}
\label{fig:main} \vspace{+0.2cm}
\end{figure*}

\section{Affectron}
In this section, we introduce Affectron, which fine-tunes a speech backbone pre-trained on purely verbal speech \cite{peng2024voicecraft} using affectively aligned NV augmentation constructed from an open-source decoupled corpus \cite{richter24_interspeech}. Figure \ref{fig:main} illustrates the training-time augmentation and fine-tuning framework of Affectron, while the detailed components are described in the following subsections.

\subsection{Emotion-Driven Top-\texorpdfstring{$K$}{K} NV Matching}
\label{sec:EDNM}
We design the emotion-driven top-$K$ NV matching module to ensure affective consistency while preserving diversity among NV inputs. Given a verbal utterance $u$ and its corresponding speaker label $s$, all NV candidates $NV^{\text{all}}_s$ associated with the speaker are retrieved. The emotional similarity between each candidate $nv_n \in NV^{\text{all}}_s$ and the utterance $u$ is calculated using Emotion2Vec embeddings \cite{ma2023emotion2vec}. The top-$K$ candidates form an index set $\mathcal{K}$, and the corresponding NV subset is denoted as $NV^{\mathcal{K}}_s$. To facilitate probability-based selection among emotionally similar candidates, the top-$K$ similarity scores are normalized using a temperature-scaled softmax distribution \cite{shazeer2017outrageously, fan-etal-2018-hierarchical, fedus2022switch}:
\begin{equation}
p(nv_k \mid u) = \frac{\exp(\mathrm{CS}(\mathbf{e}^u, \mathbf{e}^{nv}_k) / \tau)}
{\sum_{m \in \mathcal{K}} \exp(\mathrm{CS}(\mathbf{e}^u, \mathbf{e}^{nv}_m) / \tau)},
\end{equation}
where $\mathbf{e}^u$ and $\mathbf{e}^{nv}_k$ denote the emotion embeddings of $u$ and the $k$-th NV candidate $nv_k \in NV^{\mathcal{K}}_s$, $\mathrm{CS}(\cdot, \cdot)$ denotes the cosine similarity, and $\tau$ is the temperature parameter. To handle cases where NVs occur multiple times within an utterance, we sample up to two NVs from the selection distribution:
\begin{equation}
NV^{*}_s = \{S^{nv}_j \mid S^{nv}_j \sim p(nv_k \mid u),\, j \le 2 \},
\end{equation}
where $NV^{*}_s$ denotes the set of NV candidates $S^{nv}_j$ sampled from the probability distribution $p(nv_k \mid u)$ over the top-$K$ candidates. Additional analyses of alternative embedding choices for NV matching and the validity of pseudo-label-based NV--emotion alignment are provided in Appendices~\ref{appendix:embedding_comparison} and~\ref{appendix:pseudolabel}, respectively.

\subsection{Emotion-Aware Top-\texorpdfstring{$K$}{K} Routing}
\label{sec:EAR}
The emotion-aware top-$K$ routing module determines contextually appropriate locations for NV insertion. Initially, word-level segments are extracted from each verbal utterance $u$ using the Montreal Forced Aligner \cite{mcauliffe17_interspeech}. Emotional attribute pseudo-labels are assigned to each verbal segment and each NV candidate using a pre-trained emotional attribute predictor \cite{wagner2023dawn}. The extracted emotional attributes are transformed into a spherical coordinate space to quantify subtle local affective dynamics \cite{cho24_interspeech, 10965917, cho25_interspeech}. For each NV candidate $S^{nv}_j$, emotional attribute changes relative to each verbal segment $S^{vv}_i$ are measured using angular distance on the unit sphere:
\begin{multline}
\Delta(S^{nv}_j, S^{vv}_i) = \arccos( \sin\theta^{nv}_j\sin\theta^{vv}_i+ \\
\cos\theta^{nv}_j\cos\theta^{vv}_i\cos(\phi^{nv}_j-\phi^{vv}_i)).
\end{multline}

Here, $\theta$ and $\phi$ represent the elevation and azimuth angles. The affective distance $\mathrm{d}(\cdot,\cdot)$ for the $t$-th potential insertion location $\ell_t$ is computed as:
\begin{equation}
\mathrm{d}(S^{nv}_j, \ell_t) = 
\scalebox{0.7}{$
\begin{cases}
\Delta(S^{nv}_j, S^{vv}_t), & \text{if } t = 1,\\[6pt]  
\Delta(S^{nv}_j, S^{vv}_{t-1}), & \text{if } t = \mathrm{I}+1,\\[6pt]
\dfrac{\,\Delta(S^{nv}_j, S^{vv}_{t-1}) + \Delta(S^{nv}_j, S^{vv}_t)\,}{\;2\;} , & \text{otherwise.}
\end{cases}
$}
\end{equation}

Here, $\mathrm{I}$ denotes the total number of verbal segments in the utterance. After computing distances for all candidate insertion locations, the top-$K$ locations with the smallest distances are selected to define the index set $\mathcal{K}$. Subsequently, the negative distances are transformed into a temperature-scaled softmax distribution \cite{shazeer2017outrageously, fan-etal-2018-hierarchical, fedus2022switch}:
\begin{equation}
p(\ell_k|S^{nv}_j) = \frac{\exp(-\mathrm{d}(S^{nv}_j, \ell_k) / \tau)} {\sum_{m \in \mathcal{K}} \exp(-\mathrm{d}(S^{nv}_j, \ell_m) / \tau)},
\end{equation}
where $\ell_k$ denotes the $k$-th candidate insertion location and $\tau$ is the temperature parameter. The insertion location $\ell^{*}_{j}$ for the NV candidate $S^{nv}_j$ is determined by sampling from the selection distribution:
\begin{equation}
\ell^{*}_{j} \sim p(\ell_k|S^{nv}_j), \quad k \in \mathcal{K}.
\end{equation}

Detailed formulations of the spherical coordinate transformation and the routing computation procedure are provided in Appendix \ref{appendix:ear_details}.

\subsection{NV Structural Masking and Delayed Stacking}
Building on the causal masking strategy \cite{peng2024voicecraft}, we propose an NV structural masking scheme that applies this mechanism to NV codec tokens. The codec sequence is initially rearranged according to the selected insertion locations determined by emotion-aware routing. For clarity, consider a token sequence composed of a verbal span $X^{vv}=(x^{vv}_1,\dots,x^{vv}_6)$ and two NV spans $X^{nv_1}=(x^{nv_1}_{1},x^{nv_1}_{2})$ and $X^{nv_2}=(x^{nv_2}_{1},x^{nv_2}_{2})$. This sequence is rearranged into $X^{\text{re}}=(X^{vv}_1;X^{nv_1};X^{vv}_2;X^{nv_2})$, where ``$;$'' denotes concatenation, $X^{vv}_{1}=(x^{vv}_{1}, \dots , x^{vv}_{3})$, and $X^{vv}_{2}=(x^{vv}_{4}, \dots , x^{vv}_{6})$. To enable emotion-conditioned infilling based on verbal context, one NV span is randomly sampled, and a masked span is constructed around it. The masked span length is sampled as $l \sim \mathrm{Uniform}(1, L)$, which allows the span to optionally include adjacent verbal tokens. For instance, if the masked span length is 4 and the masking is applied around $X^{nv_1}$, a masked span such as $\left<MASK_{1}\right> = (x^{vv}_3, x^{nv_1}_1, x^{nv_1}_2, x^{vv}_4)$ can be selected. For each masked span $\left<MASK_{n}\right>$, the original tokens are relocated to the end of $X^{\text{re}}$, with a mask token $M_{n}$ inserted immediately in front of the span. Finally, delayed stacking \cite{copet2023simple} is applied to facilitate efficient AR modeling across parallel codebook streams, resulting in the final relocated token sequence $Z$.

\subsection{Modeling with Aligned NV-Augmented Tokens}
The Transformer decoder employs a GPT-style architecture to model the relocated codec token sequence $Z$, in which nonverbal and verbal segments are jointly aligned and augmented with NV tokens. To enable AR generation, the model is conditioned on the rearranged speech transcription $T_{re} = [t^{vv}_1;t^{vv}_2;\left<t^{nv}_1\right>;t^{vv}_3;t^{vv}_4;\left<t^{nv}_2\right>]$, constructed using the optimal insertion location $\ell^{*}_{j}$ identified by the emotion-aware routing module. The model is optimized with the standard language modeling objective, applying cross-entropy loss across all tokens \cite{peng2024voicecraft}.

\subsection{Training and Inference Workflow}
During training, Affectron constructs NV-augmented utterances by applying emotion-driven top-$K$ NV matching and emotion-aware top-$K$ routing to verbal utterances and candidate NV segments. The resulting NV-augmented transcripts and rearranged codec token sequences are then used to fine-tune the VoiceCraft backbone \cite{peng2024voicecraft} with an AR codec language modeling objective. During inference, in contrast, the model directly generates speech from an NV-tagged transcript and a reference utterance that provides the target speaker identity and emotional condition, without applying matching or routing.

\section{Experiments}
\subsection{Dataset}
The EARS dataset \cite{richter24_interspeech} was utilized for fine-tuning. It consists of approximately 100 hours of clean speech recorded under anechoic conditions from 107 speakers, encompassing diverse reading styles and emotions. In addition to verbal speech, the corpus contains separately recorded NVs across 15 types, totaling approximately four hours. Each audio file contains either multiple sentences of verbal speech or multiple stylistic variations of a single NV type, such as fillers (e.g., um, oh). Whisper \cite{radford2023robust} was employed to segment multi-sentence verbal recordings into individual utterances. Silence detection based on simple acoustic cues was applied to divide NV files containing multiple stylistic variations into individual NV events. The seen-speaker test set was constructed by sampling one random utterance for each combination of speaker and emotion, resulting in approximately 2,200 samples. To evaluate zero-shot generalization, four speakers (p001, p004 for males; p002, p003 for females) were held out as unseen speakers. An unseen-speaker test set was additionally constructed by sampling multiple random utterances for each combination of speaker and emotion, yielding approximately 400 evaluation samples.

NonverbalTTS \cite{borisov2025nonverbaltts} is a 17-hour open-access English speech corpus with aligned text annotations for 10 NV types. This dataset was used to evaluate NV locations and types. The test set includes approximately 1,000 sentences, each containing a single NV. The onset locations of NV events were labeled using FlexSED \cite{hai2025flexsed}, following prior work \cite{borisov2025nonverbaltts}.

\subsection{Implementation Details}
We initialized our model using the 330M-parameter VoiceCraft checkpoint\footnote{\url{https://huggingface.co/pyp1/VoiceCraft/tree/main}} \cite{peng2024voicecraft}, which was pre-trained on purely verbal speech \cite{chen21o_interspeech}, and employed EnCodec \cite{defossez2022high} as the speech tokenizer. For fine-tuning, we used the AdamW optimizer with a learning rate of $1\times10^{-5}$, a batch size of 100 via gradient accumulation, trained for 50,000 steps. Training was conducted on four NVIDIA RTX A6000 GPUs over a period of five days. During training, the number of masked spans was sampled from a truncated $\mathrm{Poisson}(1)$ distribution in $[1,3]$, with span lengths drawn from $\mathrm{Uniform}(1,600)$. For both the emotion-driven NV matching and emotion-aware routing modules, we set the top-$K$ values to 10 and 5, and applied a temperature parameter of $\tau = 0.7$.

\begin{figure}[!t]
  \centering
\centerline{\includegraphics[width=0.5\textwidth]{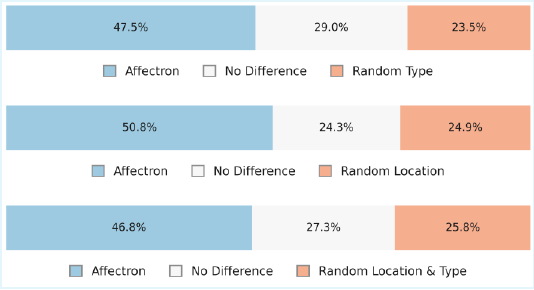}}\vspace{-0.25 cm}
\caption{AB preference test comparing the proposed NV augmentation with rule-guided randomized strategies following CapSpeech \cite{wang2025capspeech}.
}
\label{fig:abtest} \vspace{0.0cm}
\end{figure}

\begin{table*}[!ht]
    \centering 
        \resizebox{1.00\textwidth}{!}{
    \begin{tabular}{l|c|>{\centering\arraybackslash}p{1.1cm}>{\centering\arraybackslash}p{1.1cm}>{\centering\arraybackslash}p{1.1cm}|cccc|ccc}
        \toprule
        \multirow{2}{*}{\textbf{Method}} & \multirow{2}{*}{\textbf{DA}} & \multirow{2}{*}{\textbf{EDNM}} & \multirow{2}{*}{\textbf{EAR}} & \multirow{2}{*}{\textbf{NSM}} & \multicolumn{4}{c|}{\textbf{Nonverbal Metrics}} & \multicolumn{3}{c}{\textbf{Verbal Metrics}} \\
         & & & & & \textbf{NV-Acc} ($\uparrow$) & \textbf{NV-Sim} ($\uparrow$) & \textbf{EECS} ($\uparrow$) & \textbf{SECS} ($\uparrow$) & \textbf{WER} ($\downarrow$) & \textbf{EECS} ($\uparrow$) & \textbf{SECS} ($\uparrow$) \\
        \midrule
        \midrule
            \multicolumn{12}{c}{\textsc{\textcolor{gray!60}{\large Seen Speakers}}} \\
            \midrule
            Augmented GT & - & - & - & - & 85.96 & - & 0.5796 & 0.9231 & 1.13 & 0.6186 & 0.9163 \\ 
        \midrule
            VoiceCraft \cite{peng2024voicecraft} & \xmark $\cellcolor{gray!25}$ & \xmark $\cellcolor{gray!25}$ & \xmark $\cellcolor{gray!25}$ & \xmark $\cellcolor{gray!25}$ & 10.49 & 0.5898 & \textbf{0.6149} & \underline{0.8950} & 9.05 & \underline{0.6212} & \underline{0.8927} \\ 
        \midrule
            \multirow{4}{9em}{Affectron (Proposed)} & \cmark & \xmark $\cellcolor{gray!25}$ & \xmark $\cellcolor{gray!25}$ & \xmark $\cellcolor{gray!25}$ & \textbf{58.78} & 0.5988 & 0.5455 & \textbf{0.8927} & \underline{6.06} & 0.6190 & \textbf{0.8950} \\ 
            & \cmark & \cmark & \xmark $\cellcolor{gray!25}$ & \xmark $\cellcolor{gray!25}$ & 35.83 & 0.6085 & 0.5648 & 0.8897 & \textbf{6.02} & 0.6126 & 0.8892 \\ 
            & \cmark & \cmark & \cmark & \xmark $\cellcolor{gray!25}$ & 32.93 & \underline{0.6090} & 0.5707 & 0.8915 & 7.52 & 0.6211 & 0.8889 \\ 
            & \cmark & \cmark & \cmark & \cmark & \underline{37.75} & \textbf{0.6118} & \underline{0.5748} & 0.8906 & 6.59 & \textbf{0.6216} & 0.8886 \\ 
        \midrule
        \midrule
            \multicolumn{12}{c}{\textsc{\textcolor{gray!60}{\large Unseen Speakers}}} \\
            \midrule
            Augmented GT & - & - & - & - & 89.29 & - & 0.5487 & 0.9092 & 2.93 & 0.6733 & 0.8995 \\ 
        \midrule
            VoiceCraft \cite{peng2024voicecraft} & \xmark $\cellcolor{gray!25}$ & \xmark $\cellcolor{gray!25}$ & \xmark $\cellcolor{gray!25}$ & \xmark $\cellcolor{gray!25}$ & 11.90 & 0.4766 & \underline{0.5479} & \underline{0.8755} & 10.50 & \underline{0.5582} & \textbf{0.8690} \\ 
        \midrule
            \multirow{4}{9em}{Affectron (Proposed)} & \cmark & \xmark $\cellcolor{gray!25}$ & \xmark $\cellcolor{gray!25}$ & \xmark $\cellcolor{gray!25}$ & \textbf{52.38} & 0.5186 & 0.5012 & \textbf{0.8757} & 9.48 & 0.5472 & 0.8657 \\ 
            & \cmark & \cmark & \xmark $\cellcolor{gray!25}$ & \xmark $\cellcolor{gray!25}$ & 26.19 & 0.5127 & 0.5252 & 0.8694 & \underline{8.82} & 0.5580 & \underline{0.8687} \\ 
            & \cmark & \cmark & \cmark & \xmark $\cellcolor{gray!25}$ & 33.33 & \underline{0.5230} & 0.5281 & 0.8676 & 10.21 & 0.5547 & 0.8644 \\ 
            & \cmark & \cmark & \cmark & \cmark & \underline{36.90} & \textbf{0.5427} & \textbf{0.5506} & 0.8686 & \textbf{8.31} & \textbf{0.5591} & 0.8630 \\ 
        \bottomrule
    \end{tabular}
    }
    \caption{Experimental results of the proposed method for both seen and unseen speakers. DA, EDNM, EAR, and NSM denote data augmentation, emotion-driven top-$K$ NV matching, emotion-aware top-$K$ routing, and NV structural masking, respectively. Augmented GT applies our NV augmentation to the ground truth.}
    \label{tab:main}\vspace{-0.2cm}
\end{table*}

\subsection{Evaluation Metrics}
We evaluated our model using both subjective and objective metrics to assess the quality of synthesized speech. For subjective evaluation, we conducted AB preference tests and mean opinion score (MOS) assessments for NV-type naturalness (NTN-MOS) and NV-context emotional congruence (NEC-MOS). For objective evaluation, we measured verbal and nonverbal speaker embedding cosine similarity (V/NV-SECS), verbal and nonverbal emotion embedding cosine similarity (V/NV-EECS), nonverbal classification accuracy (NV-Acc), nonverbal similarity (NV-Sim), and word error rate (WER). Additionally, we evaluated NV-type and location prediction using top-$K$ accuracy (Acc@$K$), Jensen-Shannon Divergence (JSD), and Hellinger Distance (HD). Detailed definitions for metrics are provided in Appendix \ref{appendix:metrics}.

\begin{figure*}[!t]
  \centering
\centerline{\includegraphics[width=0.95\textwidth]{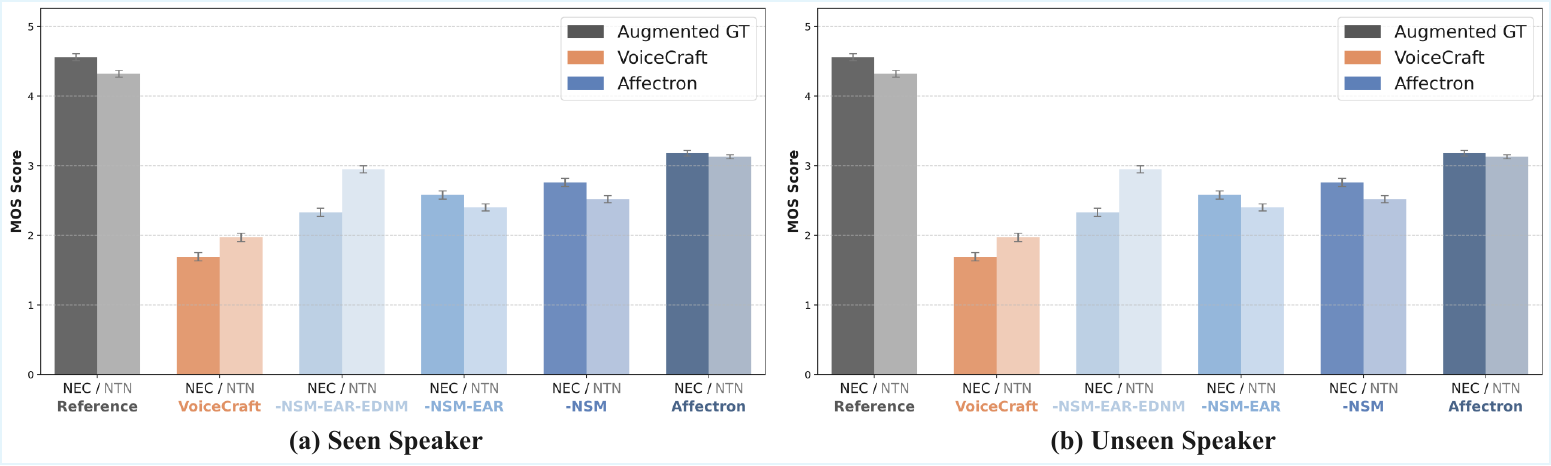}}\vspace{-0.2 cm}
\caption{Comparison of our proposed method in terms of NTN-MOS and NEC-MOS. Augmented GT applies our NV augmentation to the ground truth. Vertical lines illustrate the 95\% confidence intervals.}
\label{fig:mos} \vspace{+0.2cm}
\end{figure*}

\section{Results}
\subsection{Preference Evaluation of Augmentation}
We evaluated whether different NV augmentation methods yield perceptually natural and contextually appropriate insertions using AB preference tests on ground-truth (GT) utterances. For a fair comparison, the proposed augmentation method was replaced with three rule-guided randomized strategies, following CapSpeech \cite{wang2025capspeech}. These strategies restrict NV insertion to silent regions and randomly select NV locations and types within the permitted set. Figure \ref{fig:abtest} indicates that the proposed affect-aware NV augmentation is consistently preferred over the three rule-guided randomized strategies. These results suggest that selecting NV types and locations based on affective context leads to more natural and better-integrated NV expressions.

\subsection{Model Performance}
We performed both subjective and objective evaluations to assess the impact of each proposed module within Affectron. Experiments were conducted under two conditions: synthesis of verbal-only speech and synthesis of speech with NVs. 

As shown in Figure \ref{fig:mos} and Table \ref{tab:main}, we conducted ablation studies by removing each proposed module from the Affectron model. 1) ``w/o NSM'' discarded the NV structural masking module and reverted to baseline causal masking \cite{peng2024voicecraft}. Unlike random masking, the proposed structure leverages both past and future affective context from verbal speech during generation, enhancing the naturalness and expressiveness of the NV synthesis. 2) ``w/o EAR'' removed the emotion-aware top-$K$ routing module, inserting NVs at rule-guided random locations \cite{wang2025capspeech}. Aligning insertion locations with the progression of affective change enables the full model to achieve a more emotionally expressive integration of NV and verbal components. 3) ``w/o EDNM'' removed the emotion-driven top-$K$ NV matching module, resulting in random pairing of NVs with verbal segments from the same speaker \cite{wang2025capspeech}. Random matching increased NV diversity, resulting in higher NV-Acc and NTN-MOS, but a lack of emotional congruence with verbal speech significantly reduced EECS. 4) ``w/o DA'' removed the NV-augmented training process, resulting in separate training of verbal and NV data. Without augmentation, the model overfitted to the emotional information present in each input, resulting in higher EECS while degrading NV-related performance. 

Overall, these results demonstrate that Affectron enhances NV expressiveness and diversity while maintaining natural speech continuity. Additional comparisons with NV-capable zero-shot TTS models, analyses of speaker entanglement under cross-speaker NV mixing, and top-$K$ ablation results are provided in Appendices~\ref{appendix:NV-TTS}, \ref{appendix:speaker_mix}, and \ref{appendix:topk}, respectively.

\begin{figure}[!t]
  \centering
\centerline{\includegraphics[width=0.5\textwidth]{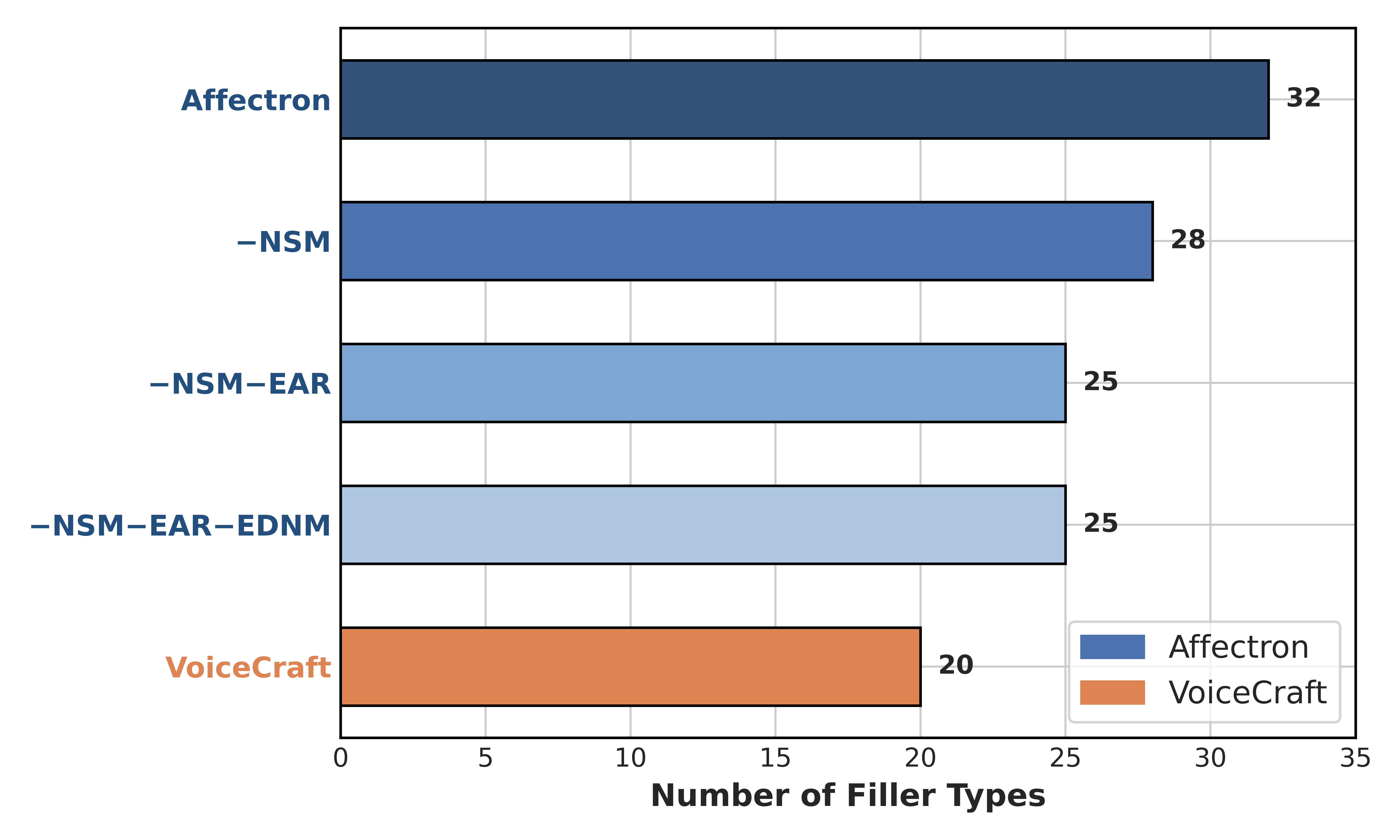}}\vspace{-0.2 cm}
\caption{Comparison of the diversity of generated fine-grained filler variations across models.}
\label{fig:fillercount} \vspace{+0.2cm}
\end{figure}

\begin{table*}[t!]
\centering
\resizebox{1.00\textwidth}{!}{
  \begin{tabular}{l|cc|cc|cc|cc|cc}
    \toprule
    \multirow{2}{*}{\textbf{Method}} & \multicolumn{2}{c|}{\textbf{Acc@1} ($\uparrow$)} & \multicolumn{2}{c|}{\textbf{Acc@3} ($\uparrow$)} & \multicolumn{2}{c|}{\textbf{Acc@5} ($\uparrow$)} & \multicolumn{2}{c|}{\textbf{\text{JSD} ($\downarrow$)}} & \multicolumn{2}{c}{\textbf{\text{HD} ($\downarrow$)}} \\
    \cmidrule{2-11} 
     & Location & Type & Location & Type & Location & Type & Location & Type & Location & Type \\
    \midrule
    Qwen 2.5-7B \cite{hui2024qwen2}   & 7.48 & \underline{25.65} & 14.13 & 47.03 & 25.30 & 69.39 & 0.1425 & 0.2139 & 0.3970 & 0.5027 \\
    LLaMA 3.1-8B \cite{dubey2024llama}  & 5.46 & 16.03 & 18.76 & 49.88 & 21.85 & \underline{77.32} & 0.4399 & 0.3020 & 0.3884 & 0.5741 \\
    GPT-oss-20B \cite{agarwal2025gpt} & \textbf{26.84} & 16.98 & \textbf{50.24} & \underline{51.19} & \textbf{62.47} & 72.68 & \underline{0.1278} & \underline{0.1130} & 0.4328 & \underline{0.3523} \\
    Vicuna-7B \cite{chiang2023vicuna} & \underline{13.78} & 11.28 & 23.63 & 41.57 & 39.07 & 69.48 & 0.1535 & 0.2431 & \underline{0.3847} & 0.5176 \\
    \midrule
    Affectron-330M (Proposed) & 12.59 & \textbf{75.77} & \underline{29.69} & \textbf{85.99} & \underline{44.06} & \textbf{91.69} & \textbf{0.0523} & \textbf{0.0051} & \textbf{0.2414} & \textbf{0.0723} \\
    \bottomrule
  \end{tabular}}
\caption{Comparison of zero-shot NV type and location prediction across LLMs and the Affectron model. Prediction accuracy is measured using top-$K$ accuracy (Acc@1/3/5) and distributional alignment is assessed using Jensen-Shannon Divergence (JSD) and Hellinger Distance (HD).}
\label{tab:llm_comparision}\vspace{-0.2cm}
\end{table*}

\subsection{Analysis of Generated Filler Diversity}
\label{result:filler_diversity}
To further investigate the expressive range of NVs, the number of distinct filler variations synthesized by each model when provided with the same NV tag was evaluated. A filler tag $\langle \text{filler} \rangle$ was inserted at affectively appropriate locations in the test set, and the realized filler type was extracted from each generated utterance, following \cite{kim-etal-2025-fillerspeech}. As shown in Figure \ref{fig:fillercount}, the number of distinct filler variations produced by the baseline model was compared with those produced by the ablated variants of Affectron. VoiceCraft \cite{peng2024voicecraft} generated only a limited subset of filler variations, and removing any of the proposed modules progressively reduced filler diversity. In contrast, Affectron produced a wider variety of fine-grained filler NV realizations, suggesting that the NV-augmented training process promotes richer NV expression. A detailed analysis of NV categories and their distributional patterns is provided in Appendix~\ref{appendix:nv_analysis}.

\subsection{Comparison with LLM Baselines on NV Type and Location Prediction}
\label{result:NV_predict}
Recent studies on NV-aligned emotional corpora \cite{xin2024jvnv} and filler placement \cite{kim-etal-2025-fillerspeech} have explored large language model (LLM)-based approaches. LLMs are capable of partially capturing discourse-level regularities, including those associated with NV expression patterns. Building on these findings, the proposed type and location prediction methods were evaluated against several LLMs without task-specific training. The evaluated LLMs include Qwen 2.5-7B \cite{hui2024qwen2}, LLaMA 3.1-8B \cite{dubey2024llama}, Vicuna-7B \cite{chiang2023vicuna}, and GPT-oss-20B \cite{agarwal2025gpt}, which represent a range of model sizes and conversational capabilities. Both NV type and location prediction were evaluated using the NonverbalTTS dataset \cite{borisov2025nonverbaltts}, which offers high-quality NV annotations suitable for model comparison. Additional details regarding the LLM baselines and prompting procedures are available in Appendix \ref{appendix:LLM}.

As shown in Table \ref{tab:llm_comparision}, Affectron achieved the highest type prediction accuracy as well as the lowest JSD and HD. These results indicate that emotion-driven top-$K$ NV matching effectively captures affective category priors that align with actual NV distributions. In terms of location prediction, GPT-oss-20B attained the highest top-$K$ accuracy. Conversely, Affectron achieved the lowest JSD and HD, demonstrating closer alignment with empirical positional distributions through its emotion-aware top-$K$ routing. Although text-only LLMs can implicitly model discourse structure, the absence of prosodic and emotional cues limits their ability to distinguish fine-grained affective nuances in NV usage. Overall, these results suggest that while text-only LLMs can capture coarse discourse structure, explicit emotion-aware modeling is necessary to achieve well-aligned NV type and location prediction. Since this comparison focuses on text-only LLM baselines motivated by prior text-driven augmentation studies, we additionally report comparisons with multimodal audio-capable LLMs in Appendix~\ref{appendix:multimodal_llm}.

\section{Conclusion}
We introduce Affectron, an NCLM-based framework for expressive NV generation with emotion-aware control mechanisms. By leveraging a small-scale open and decoupled corpus, the framework augments data through emotion-driven top-$K$ matching and emotion-aware top-$K$ routing. This approach expands the distribution of NV types and locations while preserving the naturalness of verbal content. In contrast to conventional NCLM-based TTS systems, Affectron incorporates NV structural masking, enabling smoother transitions and preserving contextual coherence. Across subjective and objective evaluations, the proposed approach improves NV realism and diversity without degrading the naturalness of verbal speech. Overall, Affectron offers a practical and scalable solution for affect-aware NV synthesis using small open corpora, thereby reducing reliance on costly alignment processes and large proprietary datasets.

\section{Limitations}
Our methodology utilizes a relatively small-scale open corpus, such as the EARS dataset \cite{richter24_interspeech}, where verbal speech and NVs are recorded independently. This limitation hinders direct comparison with models trained on large-scale or proprietary in-the-wild corpora. Additionally, the separation of verbal and nonverbal data in training restricts the model’s capacity to represent overlapping verbal and nonverbal speech segments. However, accurately modeling the overlap between verbal speech and NVs remains an open challenge \cite{ludusan20_speechprosody}. Human annotators frequently disagree regarding the precise boundaries and timing of these events \cite{truong19_interspeech}. In future research, we aim to expand the NV inventory using semi-supervised mining and to enhance the modeling of overlapping verbal and nonverbal segments.

\section{Acknowledgments}
This work was partly supported by Institute of Information \& Communications Technology Planning \& Evaluation (IITP) under the artificial intelligence graduate school program (Korea University) (No. RS-2019-II190079) and artificial intelligence star fellowship support program to nurture the best talents (IITP-2026-RS-2025-02304828) grant funded by the Korea government (MSIT).

\bibliography{refs}

\begin{figure*}[!t]
  \centering
\centerline{\includegraphics[width=0.93\textwidth]{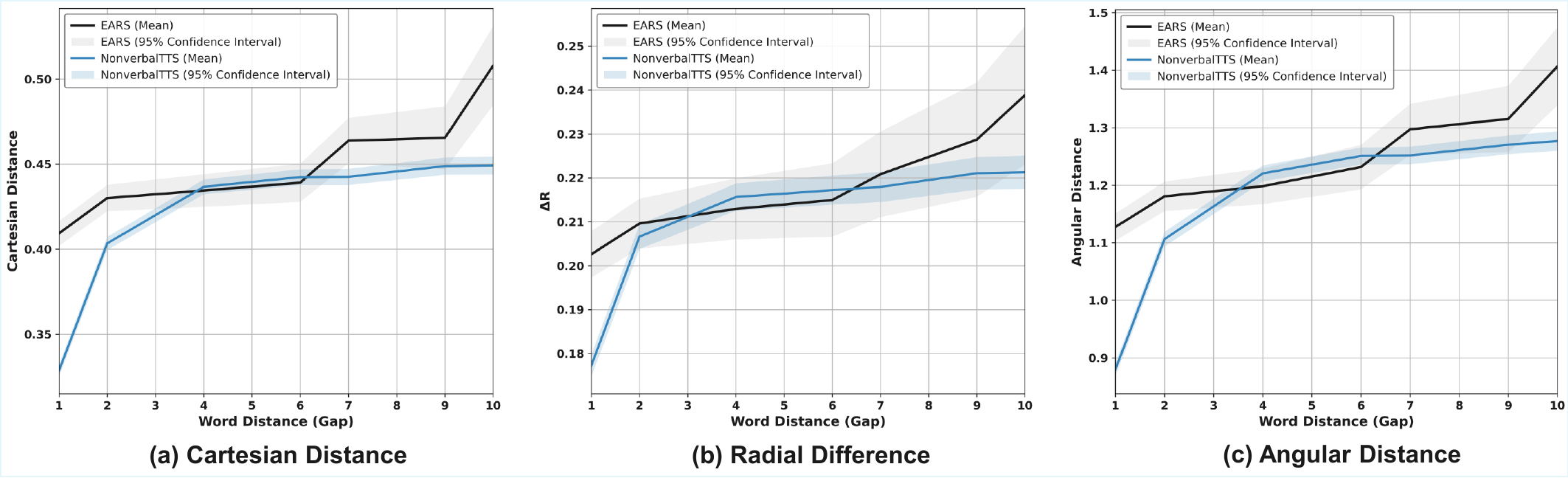}}\vspace{-0.25 cm}
\caption{Comparison of emotional attribute change patterns across temporal gaps using angular distance, Cartesian distance, and spherical radial differences.}
\label{fig:spheredistance} \vspace{0.0cm}
\end{figure*}

\newpage
\appendix

\begin{table}[t!]
\centering
\resizebox{0.5\textwidth}{!}{
\begin{tabular}{l|ccc|cc}
    \toprule
    \textbf{Method} & \textbf{Acc@1} ($\uparrow$) & \textbf{Acc@3} ($\uparrow$) & \textbf{Acc@5} ($\uparrow$) & \textbf{JSD} ($\downarrow$) & \textbf{HD} ($\downarrow$) \\
    \midrule
    Cartesian Distance  & 11.05 & 28.86 & 41.69 & 0.0568 & 0.2517  \\
    Radial Difference  & 11.88 & 28.98 & 41.45 & 0.0544 & 0.2495  \\
    Angular Distance  & \textbf{12.59} & \textbf{29.69} & \textbf{44.06} & \textbf{0.0523} & \textbf{0.2414}  \\
    \bottomrule
  \end{tabular}
}
\caption{Comparison of zero-shot NV type and location prediction across distance metrics. Prediction accuracy is measured using top-$K$ accuracy (Acc@1/3/5) and distributional alignment is assessed using Jensen-Shannon Divergence (JSD) and Hellinger Distance (HD).}
\label{table:distance}\vspace{-0.25cm}
\end{table}

\section{Analysis of Emotional Attribute Dynamics Across Distance Metrics}
\label{appendix:emotion_dynamics}
Emotional attribute changes were analyzed across multiple distance metrics to assess the robustness of the observed affective dynamics. Specifically, three distance metrics were compared to measure emotional attribute changes between pairs of speech segments separated by varying temporal gaps: (i) Cartesian distance, (ii) radial distance, and (iii) angular distance on the unit sphere, following \cite{cho24_interspeech, 10965917, cho25_interspeech}. Further details regarding emotional attribute extraction and coordinate transformation are provided in Appendix \ref{appendix:ear_details}.

\paragraph{Cartesian Distance.}
Affective changes were measured directly in Cartesian coordinates. For two speech segments $S_i$ and $S_{i+t}$ in the sequence, separated by a word-level gap $t$, the Cartesian distance is defined as:
\begin{equation}
\Delta(S_i, S_{i+t})
= \lVert \mathbf{C}_i - \mathbf{C}_{i+t} \rVert_2,
\end{equation}
where $\mathbf{C}_i = (A_i, V_i, D_i)$ denotes the emotional attribute vector of segment $S_i$. Each segment $S_i$ represents either a verbal or nonverbal segment.

\paragraph{Radial Difference.}
After transforming emotional attributes into spherical coordinates $(r, \theta, \phi)$, affective changes were measured based solely on radial magnitude. Given two speech segments $S_i$ and $S_{i+t}$ separated by a word-level gap $t$, the spherical radial difference is defined as:
\begin{equation}
\Delta(S_i, S_{i+t}) = \lvert r_i - r_{i+t} \rvert,
\end{equation}
where $r_i$ denotes the radial magnitude of segment $S_i$.

\paragraph{Angular Distance.}
To capture directional affective changes, angular distance was computed in spherical coordinate space. Given two speech segments $S_i$ and $S_{i+t}$ separated by a word-level gap $t$, the angular distance is defined as:
\begin{multline}
\Delta(S_i, S_{i+t}) = \arccos(\sin\theta_i \sin\theta_{i+t} + \\ 
\cos\theta_i \cos\theta_{i+t}\cos(\phi_i - \phi_{i+t})),
\end{multline}
where $\theta_i$ and $\phi_i$ denote the elevation and azimuth angles of segment $S_i$.

Figure \ref{fig:spheredistance} compares the emotional attribute change curves obtained using the three distance metrics for temporal gaps up to 10. For all metrics, affective differences increased as temporal gaps widened. This observation indicates that local affective states exhibit relative stability over short timescales. The influence of these differences on NV location prediction was further evaluated using each distance measure. Table \ref{table:distance} presents the top-$K$ accuracy results (Acc@$K$) and distributional alignment as measured by Jensen-Shannon Divergence (JSD) and Hellinger Distance (HD). Angular distance achieved higher top-$K$ accuracy and lower values for JSD and HD compared to the other metrics. Overall, these results confirmed that emotional attribute change patterns are consistent across distance formulations, with angular distance showing slightly more stable alignment characteristics for NV location modeling.

\begin{table}[!t]
\centering
\resizebox{0.5\textwidth}{!}{
\begin{tabular}{l|ccc|cc}
\toprule
\textbf{Method} & \textbf{Acc@1} ($\uparrow$) & \textbf{Acc@3} ($\uparrow$) & \textbf{Acc@5} ($\uparrow$) & \textbf{JSD} ($\downarrow$) & \textbf{HD} ($\downarrow$) \\
\midrule
Wav2Vec2-Base & \underline{75.15} & 85.04 & 86.94 & 0.0074 & 0.0881 \\
HuBERT-Base & 74.60 & 85.15 & \underline{88.72} & 0.0104 & 0.1041 \\
WavLM-Base & 75.10 & \underline{85.27} & 87.29 & \underline{0.0071} & \underline{0.0869} \\
CLAP & 40.93 & 58.84 & 73.43 & 0.1288 & 0.3870 \\
Emotion2Vec & \textbf{75.77} & \textbf{85.99} & \textbf{91.69} & \textbf{0.0051} & \textbf{0.0723} \\
\bottomrule
\end{tabular}
}
\caption{Comparison of alternative embedding choices for NV type prediction within the top-$K$ NV matching framework.}
\label{tab:embedding_comparison}\vspace{-0.25cm}
\end{table}

\section{Comparison with Alternative Embedding Choices for NV Matching}
\label{appendix:embedding_comparison}
To broaden the discussion of embedding choices for NV modeling, we conducted additional experiments under the NV type prediction setting described in Section~6.4 by replacing Emotion2Vec in the top-$K$ NV matching module while keeping all other components unchanged. We considered the following alternative embedding models:
\begin{itemize}
    \item \textbf{Wav2Vec2-Base}\footnote{\url{https://huggingface.co/facebook/wav2vec2-base}} \cite{baevski2020wav2vec}: a self-supervised speech representation model trained to capture general acoustic and phonetic structure from raw audio.
    \item \textbf{HuBERT-Base}\footnote{\url{https://huggingface.co/facebook/hubert-base-ls960}} \cite{hsu2021hubert}: a self-supervised speech model that learns hidden-unit representations from masked prediction over clustered acoustic targets.
    \item \textbf{WavLM-Base}\footnote{\url{https://huggingface.co/microsoft/wavlm-base}} \cite{chen2022wavlm}: a speech representation model designed to encode both content and speaker-related characteristics for speech processing tasks.
    \item \textbf{CLAP}\footnote{\url{https://huggingface.co/laion/clap-htsat-unfused}} \cite{elizalde2023clap}: a joint audio--text embedding model trained for cross-modal alignment between audio signals and natural language descriptions.
\end{itemize}
For the self-supervised speech models, NV matching was performed based on similarity between utterance and NV audio embeddings. For CLAP, NV type prediction was performed by comparing audio embeddings with textual NV label embeddings in the shared cross-modal space.

Table~\ref{tab:embedding_comparison} summarizes the results. Prosody-oriented speech representations achieved competitive performance, indicating that general speech features are helpful for NV matching. However, Emotion2Vec achieved the best overall results, including the highest Acc@1/3/5 and the lowest JSD and HD, suggesting more stable and distributionally consistent NV matching. CLAP showed substantially lower performance, possibly because its representation space is optimized for cross-modal retrieval rather than fine-grained affective alignment between speech and NV categories. Overall, these findings support the use of affect-oriented embeddings as a practical choice for emotion-aware NV matching.

\begin{table}[!t]
\centering
\resizebox{0.5\textwidth}{!}{
\begin{tabular}{l|ccccccc}
\toprule
\textbf{NV Category} & \textbf{Angry} & \textbf{Disgusted} & \textbf{Fearful} & \textbf{Happy} & \textbf{Neutral} & \textbf{Sad} & \textbf{Surprised} \\
\midrule
Agreement       & 236 & 30  & 3   & \textbf{297} & 45  & 20  & 148 \\
Anger           & \textbf{571} & 98  & 4   & 115 & 44  & 36  & 80  \\
Congratulations & 339 & 150 & 8   & 429 & 61  & 50  & \textbf{470} \\
Filler          & 72  & 89  & --  & 91  & \textbf{185} & 56  & 144 \\
Greetings       & 151 & 30  & 2   & 254 & 27  & 15  & \textbf{313} \\
Cheering        & 32  & 16  & 11  & \textbf{181} & 7   & 10  & 83  \\
Crying          & --  & 26  & 3   & 122 & 3   & \textbf{138} & 54  \\
Laughter        & 22  & 32  & 11  & \textbf{549} & 22  & 110 & 65  \\
Screaming       & 22  & 9   & \textbf{115} & 38  & 2   & 5   & 31  \\
Yelling         & 113 & 16  & 13  & \textbf{166} & 4   & 2   & 111 \\
Coughing        & 1   & \textbf{253} & 3   & 39  & 5   & 14  & 21  \\
Eating          & --  & \textbf{18} & --  & 5   & 2   & 1   & 5   \\
Sneezing        & 61  & \textbf{163} & 10  & 18  & --  & 8   & 161 \\
Throat          & 16  & \textbf{267} & 3   & 39  & 5   & 14  & 21  \\
Yawning         & 4   & 15  & --  & 10  & 1   & 9   & \textbf{43} \\
\bottomrule
\end{tabular}
}
\caption{Distribution of Emotion2Vec-based pseudo-label predictions across NV categories in the EARS dataset.}
\label{tab:pseudolabel_dist}\vspace{-0.25cm}
\end{table}

\section{Analysis of Pseudo-Label Distributions and NV--Emotion Pairing Validity}
\label{appendix:pseudolabel}
To further examine the reliability of pseudo-labels used in Affectron, we analyzed the distribution of Emotion2Vec pseudo-label across NV categories in the EARS dataset. Table~\ref{tab:pseudolabel_dist} shows that most NV categories are not associated with a single discrete emotion, but instead exhibit broad affective distributions. For example, laughter is most frequently mapped to happy, but also appears with sad and surprised predictions, which is consistent with context-dependent forms such as nervous or hollow laughter. These observations help explain why some augmented NV--emotion pairings may appear non-intuitive when viewed only at the categorical level. Affectron does not construct augmented samples based on fixed emotion--NV rules, but instead relies on embedding-level affective alignment that captures more nuanced contextual compatibility. The validity of this strategy is further supported by the strong NV type prediction and diversity results reported in Sections~\ref{result:filler_diversity} and~\ref{result:NV_predict}, suggesting that the proposed augmentation preserves meaningful NV variation beyond dominant emotion categories.

\begin{figure*}[!t]
  \centering
\centerline{\includegraphics[width=1.0\textwidth]{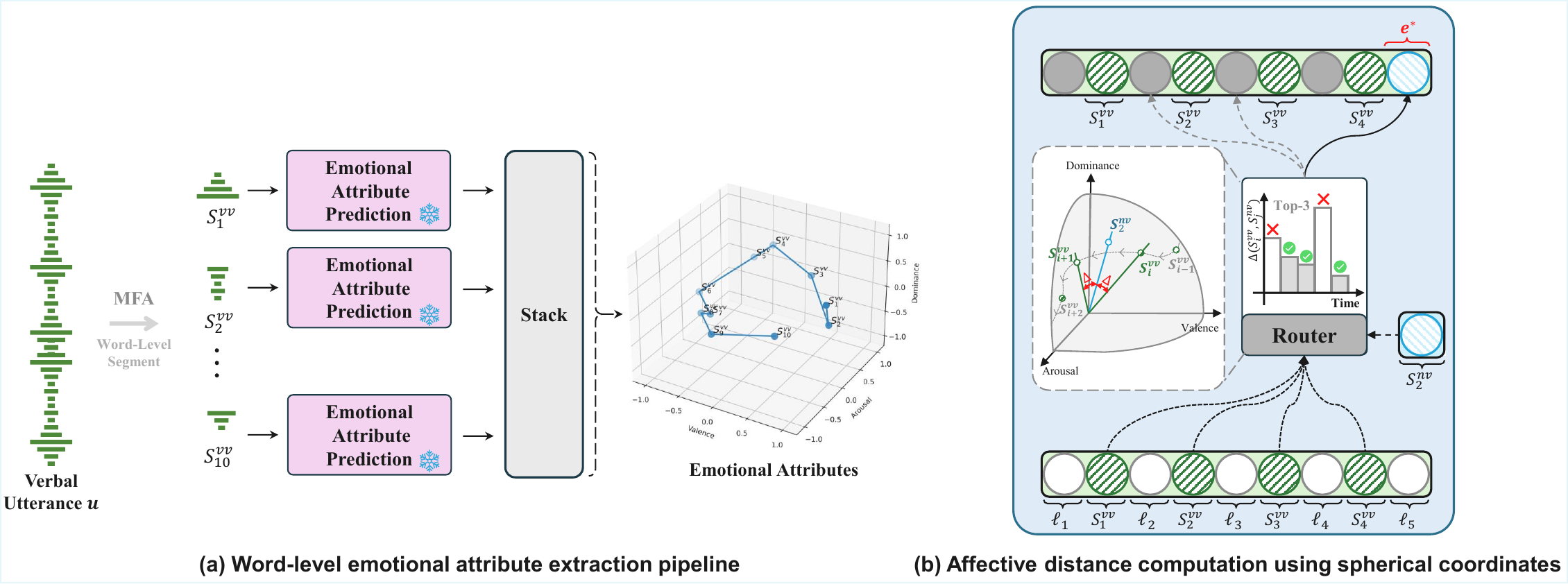}}\vspace{-0.25 cm}
\caption{Overview of the emotion-aware top-$K$ routing process. (a) Word-level emotional attributes are extracted from verbal speech using MFA and an emotional attribute predictor. (b) Affective distance computation between an NV candidate and verbal segments via angular distance in spherical coordinates.}
\label{fig:overviewear} \vspace{0.0cm}
\end{figure*}

\section{Detailed Description of Emotion-Aware Top-\texorpdfstring{$K$}{K} Routing}
\label{appendix:ear_details}
We provided a detailed description of the emotion-aware top-$K$ routing procedure introduced in Section \ref{sec:EAR}. As illustrated in Figure \ref{fig:overviewear} (a), word-level boundaries were first obtained using the Montreal Forced Aligner\footnote{\url{https://montreal-forced-aligner.readthedocs.io}} \cite{mcauliffe17_interspeech}. Word-level emotional attribute pseudo-labels for each verbal segment and each NV candidate were extracted using a pre-trained emotional attribute predictor\footnote{\url{https://huggingface.co/audeering/wav2vec2-large-robust-12-ft-emotion-msp-dim}} \cite{wagner2023dawn}. Pseudo-labels were generated in the arousal–valence–dominance space, with each dimension ranging approximately from 0 to 1 in Cartesian coordinates. First, the Cartesian coordinates were translated so that the neutral emotion center $M$ was positioned at the origin. Specifically, each emotional attribute pseudo-label $e$ was translated as follows:
\begin{equation}
e' = e - M, \quad \text{where} \quad M = \frac{1}{N_{n}} \sum_{i=1}^{N_{n}} e^{n}_{i},
\end{equation}
where $N_{n}$ denotes the total number of neutral emotion pseudo-labels $e^{n}_{i}$. The translated pseudo-labels $e'=(A,V,D)$, where $A$ denotes arousal, $V$ valence, and $D$ dominance, were then transformed into spherical coordinates $(r,\theta,\phi)$ \cite{cho24_interspeech, 10965917, cho25_interspeech} as follows:
\begin{equation}
r = \sqrt{{A}^{2} + {V}^{2} + {D}^{2}},
\end{equation}
\begin{equation}
\theta = \arccos\left(\frac{D}{r}\right), \quad
\phi = \arctan\left(\frac{V}{A}\right),
\end{equation}
where $\theta$ and $\phi$ correspond to the elevation and azimuth angles, which are used to compute affective distances during the emotion-aware top-$K$ routing procedure.

As shown in Figure \ref{fig:overviewear} (b), the affective difference between an NV candidate $S^{nv}_j$ and each verbal segment $S^{vv}_i$ was computed using the angular distance on the unit sphere. These angular distances were then aggregated to obtain an affective distance $\mathrm{d}(S^{nv}_j, \ell_t)$ for each candidate insertion index $\ell_t$. Based on these distances, the top-$K$ locations with the smallest values were selected. Finally, a temperature-scaled softmax \cite{shazeer2017outrageously, fan-etal-2018-hierarchical, fedus2022switch} over the negative distances defined the routing distribution $p(\ell_k|S^{nv}_j)$, from which the insertion index $\ell_j^{*}$ was sampled for each NV candidate.

\section{Detailed Description of Evaluation Metrics}
\label{appendix:metrics}
\subsection{Subjective Evaluation}
For subjective evaluation, we conducted AB preference tests and mean opinion score (MOS) evaluations. In AB preference tests, listeners compared pairs of utterances in which different augmentation strategies were applied to the same ground-truth (GT) speech and selected the sample exhibiting more natural and contextually appropriate NVs. For evaluation data construction, 100 utterances were randomly sampled from the test set. For MOS evaluation, we focused on two perceptual aspects of NV expressiveness: naturalness and correctness of the realized NV type (NTN-MOS), and emotional congruence of the NV with the surrounding verbal context (NEC-MOS). For evaluation data construction, a total of 150 utterances were randomly sampled from the test set, comprising 100 samples from seen speakers and 50 from unseen speakers. All MOS scores were collected using a five-point scale and were reported with 95\% confidence intervals.

\begin{figure*}[!t]
  \centering
\centerline{\includegraphics[width=1.0\textwidth]{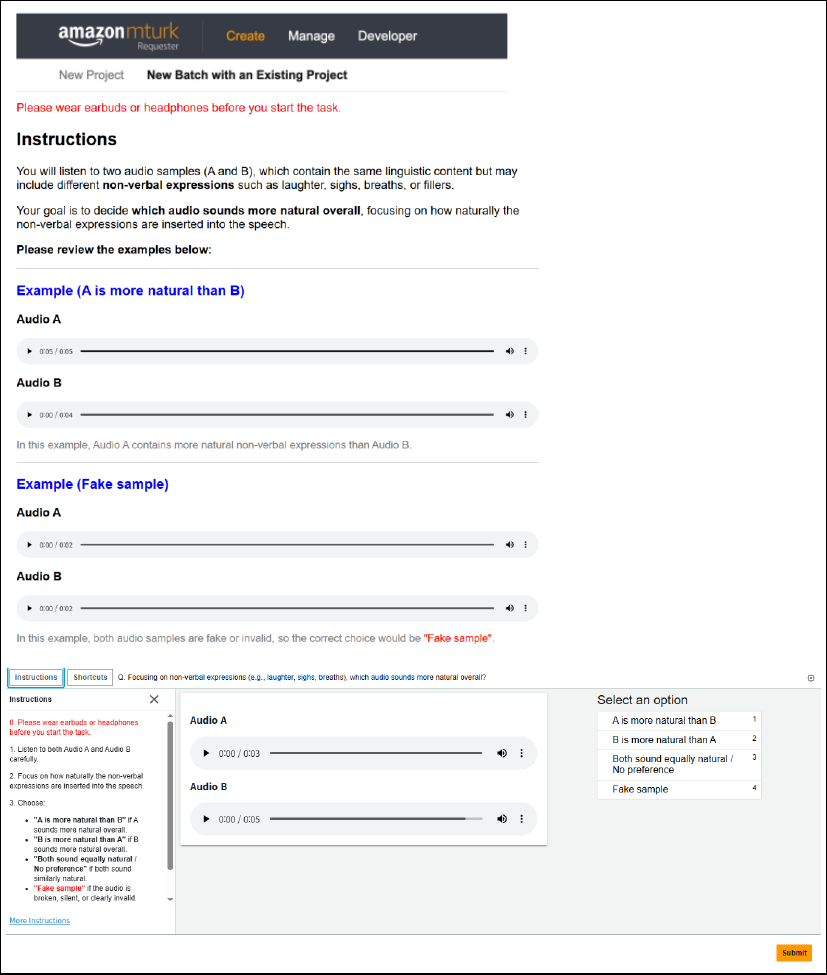}}\vspace{-0.25 cm}
\caption{Detailed information on listener requirements and the AB preference test interfaces.}
\label{Mturk_AB} \vspace{0.0cm}
\end{figure*}

\begin{figure*}[!t]
  \centering
\centerline{\includegraphics[width=1.0\textwidth]{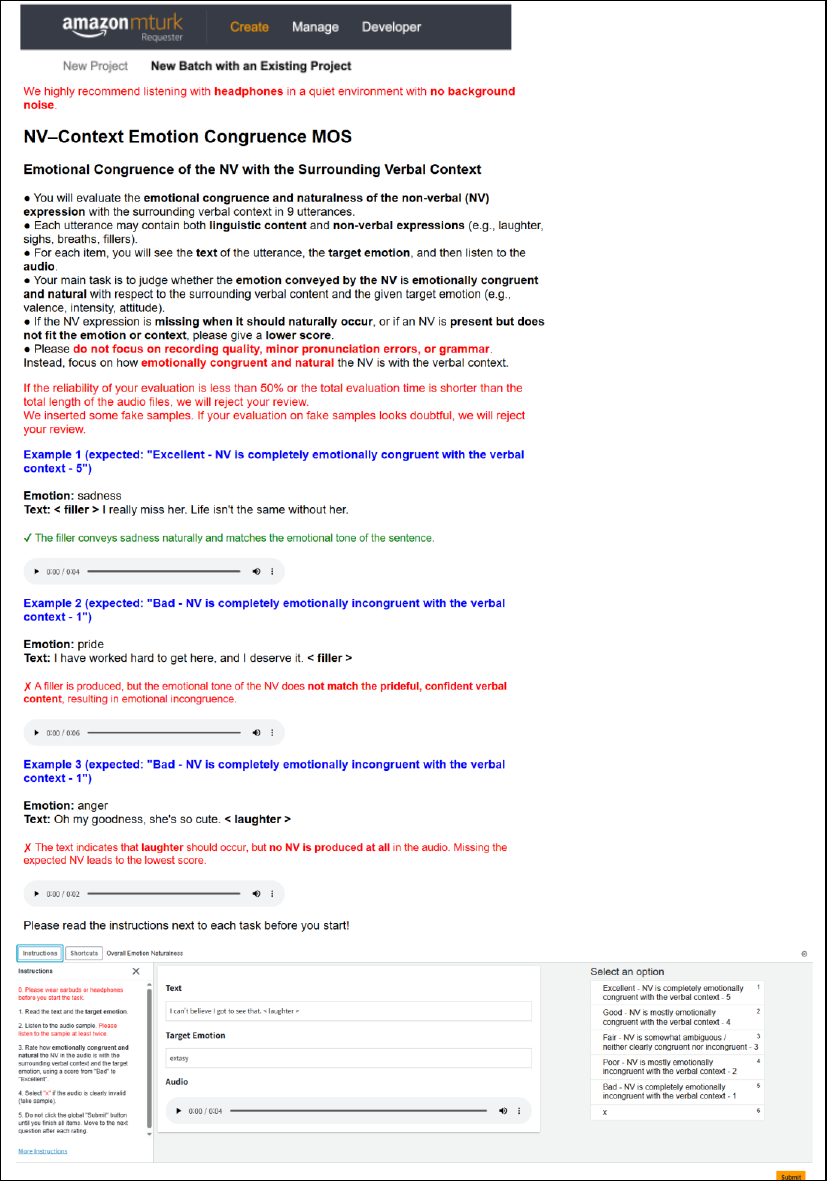}}\vspace{-0.25 cm}
\caption{Detailed information on listener requirements and NV-context emotional congruence (NEC-MOS) evaluation interfaces.}
\label{Mturk_NEC-MOS} \vspace{0.0cm}
\end{figure*}

\begin{figure*}[!t]
  \centering
\centerline{\includegraphics[width=1.0\textwidth]{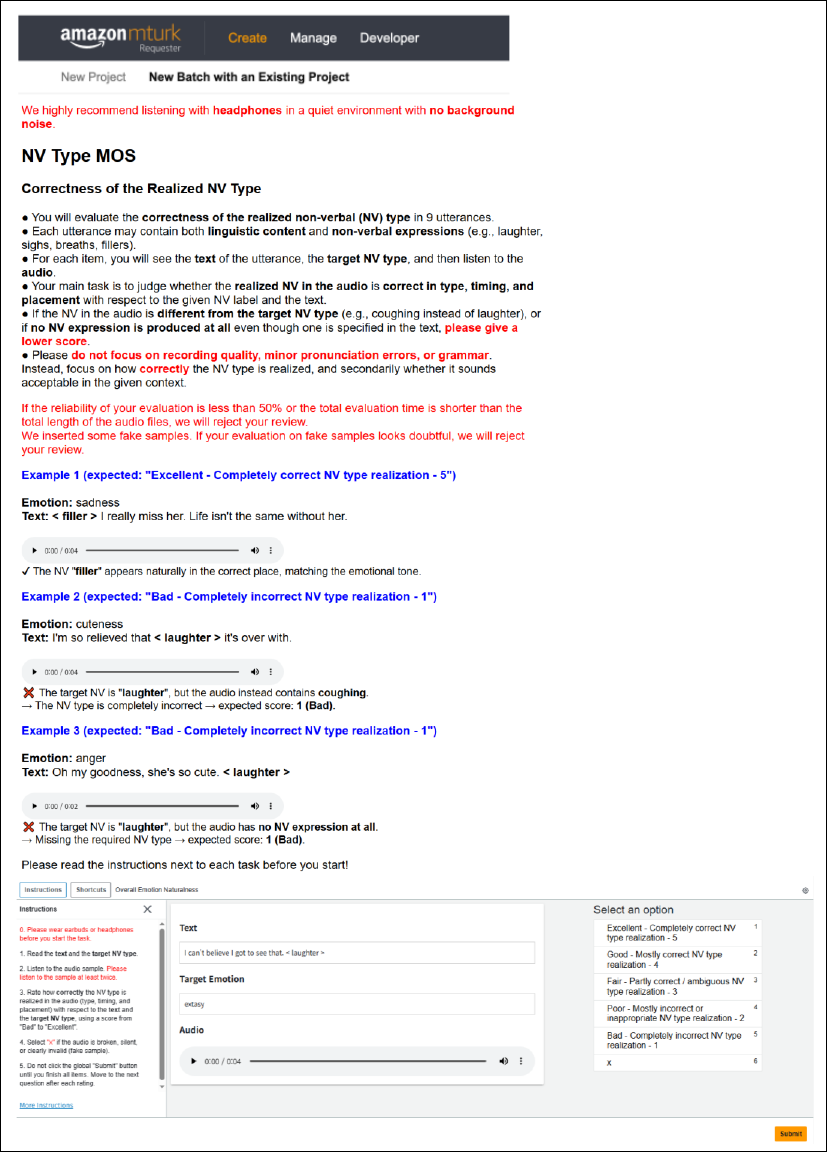}}\vspace{-0.25 cm}
\caption{Detailed information on listener requirements and NV-type naturalness (NTN-MOS) evaluation interfaces.}
\label{Mturk_NTN-MOS} \vspace{0.0cm}
\end{figure*}

We utilized crowdsourcing for these evaluations via Amazon Mechanical Turk\footnote{\url{https://www.mturk.com/}}.  A total of 20 native English speakers residing in the United States participated in each evaluation. AB preference tests and MOS-based evaluations were conducted with 20 participants each, at total costs of 60 USD and 180 USD, respectively. To ensure evaluator reliability, only workers with an approval rate of at least 98\% and more than 90 approved HITs were permitted to participate. Attention checks were implemented by inserting fake samples and verifying whether participants rated them appropriately. Evaluations from listeners who failed the attention check or spent less than half of the audio duration on the task were excluded from analysis. These measures were implemented to filter out inattentive listeners and ensure the integrity of the collected ratings. Further details on listener requirements and the evaluation interfaces for AB preference tests, NEC-MOS, and NTN-MOS are provided in Figures~\ref{Mturk_AB}, \ref{Mturk_NEC-MOS}, and \ref{Mturk_NTN-MOS}.

\subsection{Objective Evaluation}
For objective evaluation, we employed several metrics to quantify the similarity and consistency between reference and generated speech in both verbal and nonverbal domains. Experiments were conducted under two synthesis conditions: verbal-only speech synthesis and speech synthesis with NVs. Verbal metrics were computed using samples generated under the verbal-only condition, whereas nonverbal metrics were evaluated using samples synthesized with NVs.

Speaker similarity was measured using verbal and nonverbal speaker embedding cosine similarity (V/NV-SECS) using WavLM-base for speaker verification\footnote{\url{https://huggingface.co/microsoft/wavlm-base-sv}}, where cosine similarity is computed between the reference verbal speech and the generated verbal speech or NVs. Emotional expressiveness was assessed using verbal and nonverbal emotion embedding cosine similarity (V/NV-EECS) with Emotion2Vec\footnote{\url{https://github.com/ddlBoJack/emotion2vec}} \cite{ma2023emotion2vec}, which computed emotion embeddings between the reference and generated speech, following \cite{oh2024durflex}. Nonverbal classification accuracy (NV-Acc) was further evaluated using an internally trained classifier on the EARS dataset \cite{richter24_interspeech}, following the architecture of a categorical emotion recognition model \cite{goncalves2024odyssey}. Nonverbal similarity (NV-Sim) was computed as the cosine similarity between the generated NV and the GT NV selected via emotion-driven matching, using Emotion2Vec embeddings. Linguistic consistency was assessed by calculating the word error rate (WER) with the Whisper large model \cite{radford2023robust}.

Finally, we employed two complementary metrics to evaluate the accuracy of NV type and location prediction. Top-$K$ accuracy (Acc@$K$) was reported, measuring whether the GT NV type or location is included among the top-$K$ predictions. JSD \cite{lin2002divergence}, and HD \cite{beran1977minimum} were also computed between the GT and predicted distributions of NV types and locations to quantify how closely the model reproduces the occurrence patterns observed in real data.

\begin{table*}[!ht]
        \resizebox{1.0\textwidth}{!}{
    \begin{tabular}{l|cccc|ccc}
        \toprule
        \multirow{2}{*}{\textbf{Method}} & \multicolumn{4}{c|}{\textbf{Nonverbal Metrics}} & \multicolumn{3}{c}{\textbf{Verbal Metrics}} \\
         & \textbf{NV-Acc} ($\uparrow$) & \textbf{NV-Sim} ($\uparrow$) & \textbf{EECS} ($\uparrow$) & \textbf{SECS} ($\uparrow$) & \textbf{WER} ($\downarrow$) & \textbf{EECS} ($\uparrow$) & \textbf{SECS} ($\uparrow$) \\
        \midrule
            Augmented GT & 89.29 & - & 0.5487 & 0.9092 & 2.93 & 0.6733 & 0.8995 \\ 
        \midrule
            VoiceCraft-330M$^\spadesuit$ \cite{peng2024voicecraft} & - & - & - & - & 8.89 & 0.5410 & 0.8483 \\ 
            CosyVoice2-0.5B$^\spadesuit$ \cite{du2024cosyvoice} & 25.00 & 0.4097 & 0.4991 & 0.8751 & \underline{1.97} & 0.4876 & \underline{0.8739} \\ 
            Fun-CosyVoice3-0.5B$^\spadesuit$ \cite{du2025cosyvoice} & \underline{27.38} & 0.4222 & 0.4792 & \textbf{0.8952} & \textbf{1.65} & 0.4904 & \textbf{0.8916} \\ 
            Dia-1.6B$^\spadesuit$ \cite{dia} & 13.10 & \underline{0.4894} & 0.4549 & 0.8484 & 11.5 & 0.5204 & 0.8690 \\ 
        \midrule
            VoiceCraft-330M$^\clubsuit$ \cite{peng2024voicecraft} & 11.90 & 0.4766 & \underline{0.5479} & \underline{0.8755} & 10.5 & \underline{0.5582} & 0.8690 \\ 
            Affectron-330M$^\clubsuit$ (Proposed) & \textbf{36.90} & \textbf{0.5427} & \textbf{0.5506} & 0.8686 & 8.31 & \textbf{0.5591} & 0.8630 \\ 
        \bottomrule
    \end{tabular}
      }
    \centering 
        \caption{Comparison of NV-capable zero-shot TTS models on unseen speakers in the EARS dataset \cite{richter24_interspeech} zero-shot evaluation setting. $\spadesuit$ denotes a pre-trained model using the official implementation without any training or fine-tuning on the EARS dataset. $\clubsuit$ denotes a model fine-tuned on the EARS dataset based on the official pre-trained checkpoint.}
    \label{tab:NVTTS}\vspace{-0.25cm}
\end{table*}

\begin{table*}[!ht]
    \centering 
        \resizebox{0.9\textwidth}{!}{
    \begin{tabular}{l|cccc|ccc}
        \toprule
        \multirow{2}{*}{\textbf{Method}} & \multicolumn{4}{c|}{\textbf{Nonverbal Metrics}} & \multicolumn{3}{c}{\textbf{Verbal Metrics}} \\
         & \textbf{NV-Acc} ($\uparrow$) & \textbf{NV-Sim} ($\uparrow$) & \textbf{EECS} ($\uparrow$) & \textbf{SECS} ($\uparrow$) & \textbf{WER} ($\downarrow$) & \textbf{EECS} ($\uparrow$) & \textbf{SECS} ($\uparrow$) \\
        \midrule
            Augmented GT & 89.29 & - & 0.5487 & 0.9092 & 2.93 & 0.6733 & 0.8995 \\ 
        \midrule
            Cross-speaker NV mixing & 14.29 & \textbf{0.4939} & 0.4731 & 0.8571 & 9.40 & \textbf{0.5520} & 0.8604 \\ 
            Same-speaker NV mixing & \textbf{16.67} & 0.4819 & \textbf{0.4979} & \textbf{0.8675} & \textbf{8.89} & 0.5518 & \textbf{0.8660} \\ 
        \bottomrule
    \end{tabular}}
    \caption{Ablation study on cross-speaker and same-speaker NV mixing in unseen-speaker zero-shot TTS synthesis.}
    \label{tab:cross_speaker}\vspace{-0.25cm}
\end{table*}

\section{Comparison with Zero-Shot TTS Models Trained on NV-Labeled Data}
\label{appendix:NV-TTS}
Recent zero-shot TTS models have demonstrated the capability to generate NVs. However, direct comparison with existing NV-capable systems remains challenging due to differences in data availability and annotation protocols. To provide a supplementary point of reference, we additionally compared our method with publicly released NV-capable zero-shot TTS models trained on large-scale NV-labeled corpora. Specifically, we considered the following NV-capable systems:
\begin{itemize}
    \item \textbf{CosyVoice2-0.5B}\footnote{\url{https://huggingface.co/FunAudioLLM/CosyVoice2-0.5B}} \cite{du2024cosyvoice}: an improved streaming neural codec language model (NCLM) for zero-shot TTS, trained on large-scale multilingual speech data with NV annotations. The model supports 12 predefined NV categories.
    \item \textbf{Fun-CosyVoice3-0.5B}\footnote{\url{https://huggingface.co/FunAudioLLM/Fun-CosyVoice3-0.5B-2512}} \cite{du2025cosyvoice}: an improved NCLM for zero-shot multilingual TTS, extending CosyVoice2 with enhanced data scale and modeling capacity. The model supports 12 predefined NV categories.
    \item \textbf{Dia-1.6B}\footnote{\url{https://huggingface.co/nari-labs/Dia-1.6B-0626}} \cite{dia}:
    a zero-shot TTS system trained on a large NV-labeled corpus that directly generates expressive speech from transcripts. Dia can produce a broad inventory of 21 predefined NV categories.
\end{itemize}

All compared models were evaluated in a zero-shot setting using their publicly available pre-trained checkpoints. All generated audio was resampled to a uniform 16 kHz sampling rate prior to evaluation to ensure a fair comparison. Because the supported NV tag categories did not exactly match those used in this evaluation, semantically compatible NV tags were mapped to a shared representation (e.g., $\langle \text{coughing} \rangle$ to $\langle \text{cough} \rangle$, $\langle \text{screaming} \rangle$ to $\langle \text{noise} \rangle$). Accordingly, the category set used for NV classification was adapted for each model based on its supported NV inventory.

As shown in Table \ref{tab:NVTTS}, existing NV-capable zero-shot TTS models demonstrated complementary strengths across both verbal and nonverbal dimensions. CosyVoice2 and CosyVoice3 achieved strong performance on linguistic and speaker-related metrics, reflecting the advantages of large-scale supervised training and robust acoustic modeling. However, their performance on NV-related metrics was comparatively limited, suggesting that NV generation was not explicitly optimized, despite NV annotations being available during training. Dia trained on a broader inventory of NV categories, demonstrated relatively higher nonverbal similarity compared to CosyVoice-based models. Nevertheless, its NV accuracy and emotion consistency scores were comparatively lower. In contrast, Affectron outperformed baseline systems on NV-related metrics while maintaining competitive verbal quality. These findings indicate that, beyond data scale, explicitly modeling affective alignment for NV selection and placement plays a critical role in achieving coherent and expressive NV synthesis.

\begin{table*}[!ht]
    \centering 
        \resizebox{1.0\textwidth}{!}{
    \begin{tabular}{l|>{\centering\arraybackslash}p{1.1cm}>{\centering\arraybackslash}p{1.1cm}|cccc|ccc}
        \toprule
        \multirow{2}{*}{\textbf{Method}} & \multirow{2}{*}{\textbf{$K_{\text{EDNM}}$}} & \multirow{2}{*}{\textbf{$K_{\text{EAR}}$}} & \multicolumn{4}{c|}{\textbf{Nonverbal Metrics}} & \multicolumn{3}{c}{\textbf{Verbal Metrics}} \\
         & & & \textbf{NV-Acc} ($\uparrow$) & \textbf{NV-Sim} ($\uparrow$) & \textbf{EECS} ($\uparrow$) & \textbf{SECS} ($\uparrow$) & \textbf{WER} ($\downarrow$) & \textbf{EECS} ($\uparrow$) & \textbf{SECS} ($\uparrow$) \\
        \midrule
        \midrule
            \multicolumn{10}{c}{\textsc{\textcolor{gray!60}{\large Seen Speakers}}} \\
            \midrule
            Augmented GT & - & - & 85.96 & - & 0.5796 & 0.9231 & 1.13 & 0.6186 & 0.9163 \\ 
        \midrule
            \multirow{7}{9em}{Affectron (Proposed)} & 1 & 1 & 27.92 & 0.5965 & 0.5684 & 0.8887 & 8.35 & 0.6148 & 0.8885 \\ 
            & 3 & 3 & 32.18 & 0.5961 & 0.5673 & 0.8891 & 6.59 & 0.6216 & 0.8886 \\ 
            & 5 & 5 & 36.16 & 0.5970 & 0.5656 & 0.8905 & 6.19 & 0.6155 & 0.8878 \\ 
            & 10 & 5 & 37.75 & 0.6118 & 0.5748 & 0.8906 & 6.59 & 0.6216 & 0.8886 \\ 
            & 10 & 10 & 42.48 & 0.6084 & 0.5620 & 0.8900 & 6.19 & 0.6153 & 0.8899 \\ 
            & 15 & 10 & 45.95 & 0.6183 & 0.5611 & 0.8936 & 5.99 & 0.6117 & 0.8936 \\ 
            & 15 & 15 & 42.80 & 0.6106 & 0.5521 & 0.8921 & 7.06 & 0.6073 & 0.8898 \\ 
        \midrule
        \midrule
            \multicolumn{10}{c}{\textsc{\textcolor{gray!60}{\large Unseen Speakers}}} \\
            \midrule
            Augmented GT & - & - & 89.29 & - & 0.5487 & 0.9092 & 2.93 & 0.6733 & 0.8995 \\ 
        \midrule
            \multirow{7}{9em}{Affectron (Proposed)} & 1 & 1 & 27.38 & 0.5067 & 0.5261 & 0.8662 & 11.52 & 0.5683 & 0.8651 \\ 
            & 3 & 3 & 23.81 & 0.5041 & 0.5182 & 0.8699 & 9.41 & 0.5669 & 0.8690 \\ 
            & 5 & 5 & 33.33 & 0.5132 & 0.5168 & 0.8635 & 9.30 & 0.5449 & 0.8688 \\ 
            & 10 & 5 & 36.90 & 0.5427 & 0.5506 & 0.8686 & 8.31 & 0.5591 & 0.8630 \\ 
            & 10 & 10 & 34.52 & 0.5075 & 0.5342 & 0.8689 & 9.97 & 0.5670 & 0.8665 \\ 
            & 15 & 10 & 43.33 & 0.5224 & 0.5306 & 0.8728 & 13.12 & 0.5416 & 0.8665 \\ 
            & 15 & 15 & 50.00 & 0.5211 & 0.5269 & 0.8729 & 9.06 & 0.5443 & 0.8671 \\ 
        \bottomrule
    \end{tabular}}
    \caption{Grid search over the top-$K$ hyperparameters for emotion-driven NV matching (EDNM) and emotion-aware routing (EAR). Augmented GT applies our NV augmentation to the ground truth.}
    \label{tab:hyper}\vspace{-0.25cm}
\end{table*}

\section{Analysis of Speaker Entanglement and Cross-Speaker NV Mixing}
\label{appendix:speaker_mix}
To further examine the concern regarding speaker entanglement, we analyzed the effect of relaxing the same-speaker constraint used in NV augmentation. In Affectron, the same-speaker constraint was introduced to preserve speaker voice coherence and stabilize acoustic transitions between verbal speech and inserted NV segments, rather than to enforce speaker-specific NV patterns. Importantly, the candidate selection itself is still guided by affective similarity in a shared embedding space, indicating that emotional alignment remains the primary criterion during NV matching. This interpretation is also consistent with the results in Table~\ref{tab:main}, where Affectron maintains competitive NV-related performance under the unseen-speaker zero-shot setting.

To further validate this point, we conducted an additional ablation study with cross-speaker NV mixing under the same training setup for 10,000 steps and evaluated the model on unseen-speaker zero-shot TTS synthesis. As shown in Table~\ref{tab:cross_speaker}, cross-speaker mixing slightly improved NV-Sim, but did not improve affective consistency and resulted in lower speaker similarity and weaker acoustic coherence overall. In contrast, same-speaker NV mixing achieved more balanced performance across NV-related and verbal metrics. These findings suggest that the same-speaker constraint mainly serves as an acoustic stabilization strategy, while affect-driven matching remains the key mechanism enabling robust NV generalization beyond seen speakers.

\begin{figure*}[!t]
  \centering
\centerline{\includegraphics[width=0.9\textwidth]{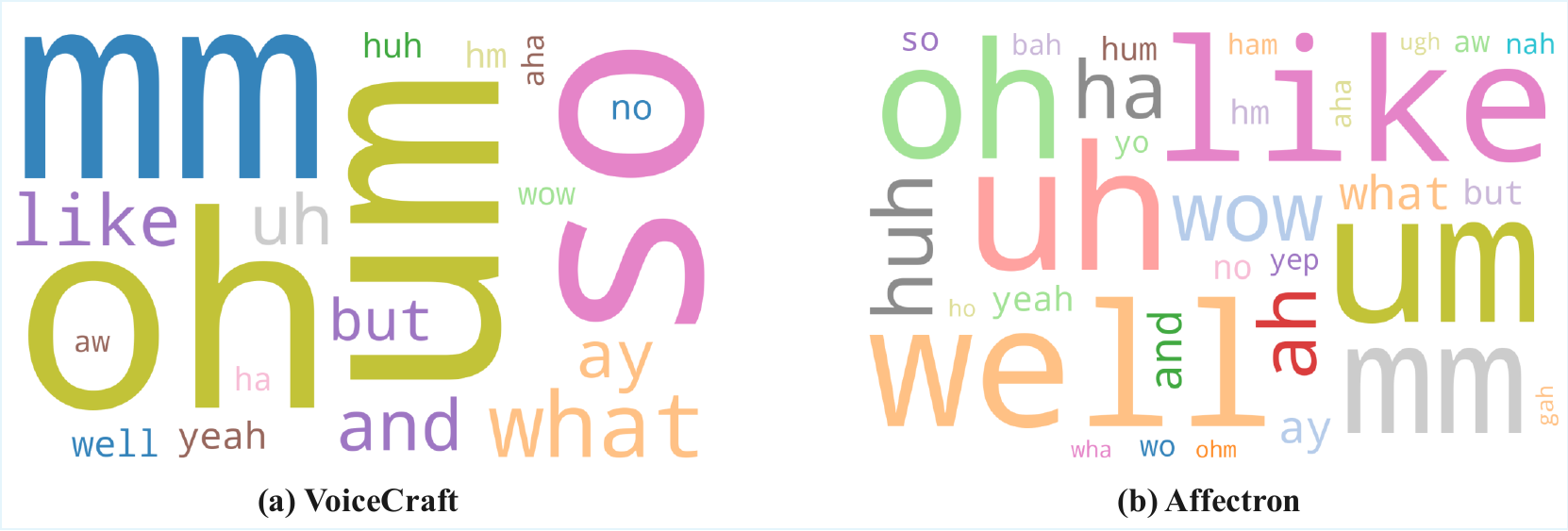}}\vspace{-0.25 cm}
\caption{Word-cloud visualization of generated fine-grained filler variations for VoiceCraft \cite{peng2024voicecraft} and Affectron. The size of each keyword corresponds to its relative frequency among the realized filler variants.}
\label{fig:wordcloud} \vspace{0cm}
\end{figure*}

\section{Ablation Analysis of Top-\texorpdfstring{$K$}{K} Hyperparameters for NV Matching and Routing}
\label{appendix:topk}
Top-$K$ selection was implemented to balance affective consistency and diversity by enabling controlled probabilistic sampling. We analyzed the sensitivity of Affectron to the top-$K$ hyperparameters used in emotion-driven NV matching (EDNM) and emotion-aware routing (EAR). Specifically, $K_{\text{EDNM}}$ and $K_{\text{EAR}}$ were varied from 1 to 15, with each configuration trained and evaluated under otherwise identical experimental settings.

As shown in Table \ref{tab:hyper}, setting $K$ to a small value resulted in both NV matching and routing becoming largely deterministic. Such restrictive configurations limited probabilistic sampling, which reduced NV diversity and decreased the overall naturalness of generated speech. Conversely, excessively large $K_{\text{EDNM}}$ values expanded the candidate set, increasing NV diversity but reducing nonverbal similarity. Similarly, overly large $K_{\text{EAR}}$ values weakened the emotional coherence of NV placement, resulting in degraded emotional expressiveness. Overall, the results indicate that intermediate top-$K$ settings offer a beneficial trade-off between NV diversity and affective coherence. Specifically, the configuration with $K_{\text{EDNM}}=10$ and $K_{\text{EAR}}=5$ consistently achieved balanced performance across both nonverbal and verbal metrics. Accordingly, this setting was adopted as a practical operating point for Affectron in all subsequent experiments.

\section{Filler NV Variations Analysis}
\label{appendix:nv_analysis}
To supplement the analysis of filler diversity, a visualization of the distribution of generated fine-grained filler NV variations was provided. As shown in Figure \ref{fig:wordcloud}, word-cloud representations that visualize the realized filler variants produced by both the VoiceCraft \cite{peng2024voicecraft} and the proposed model under identical filler tag $\langle \text{filler} \rangle$ conditions. Within the word-cloud visualization, the size of each keyword corresponds to its relative frequency among the generated filler realizations. This representation highlights both the number of distinct filler variations and their distributional balance and usage patterns.

VoiceCraft demonstrates a strong concentration on a limited set of common filler types. This pattern indicates repetitive generation behavior and limited expressive diversity. In contrast, Affectron generates a broader range of filler variants with a more balanced frequency distribution, including less frequent yet contextually appropriate realizations. Collectively, these results suggest that the proposed NV-augmented training framework encourages richer and more diverse NV expression beyond dominant filler variants.

\begin{figure}[!t]
  \centering
\centerline{\includegraphics[width=0.5\textwidth]{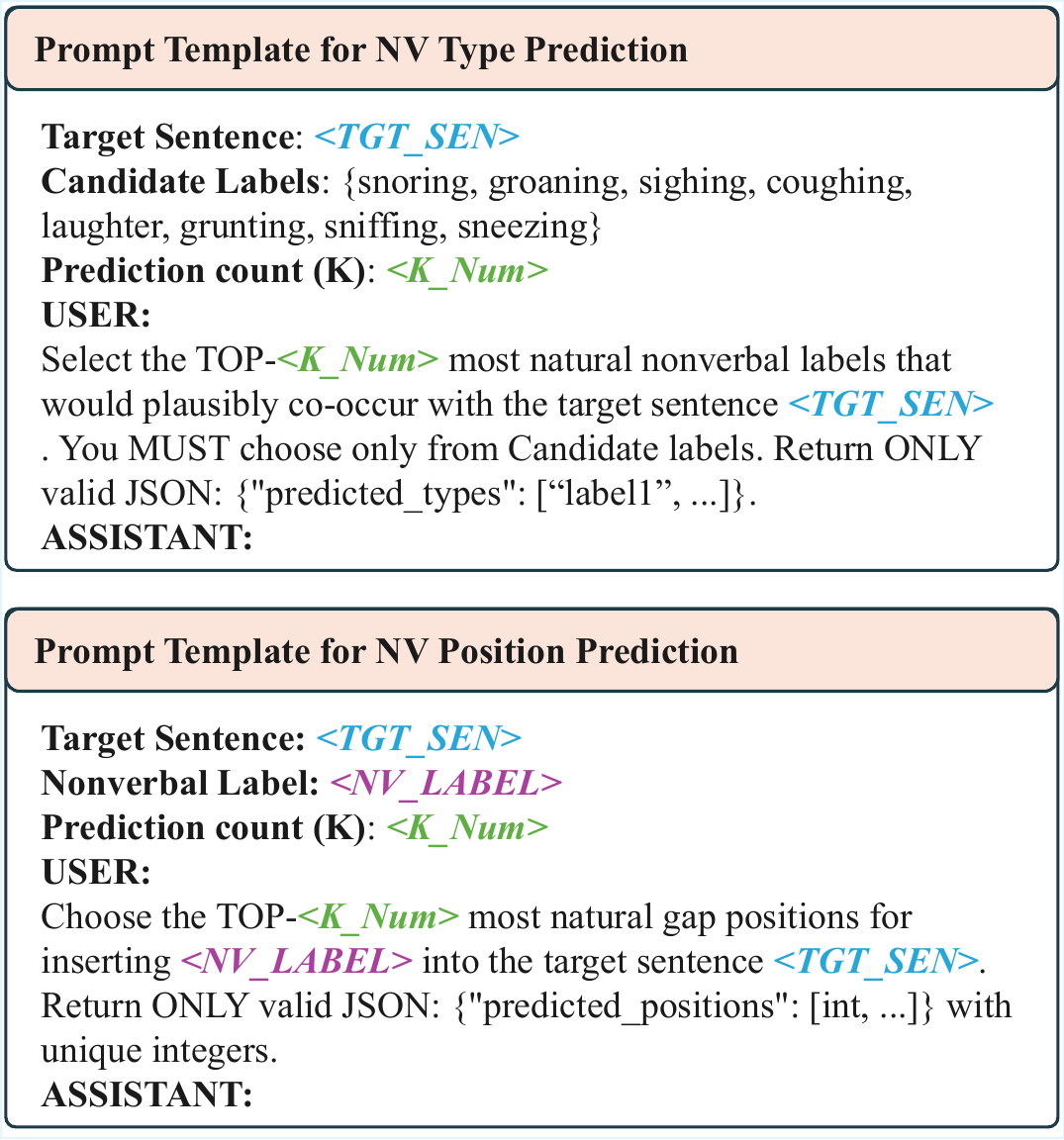}}\vspace{-0.25 cm}
\caption{Sample prompt templates for NV type and location prediction.}
\label{fig:template} \vspace{0cm}
\end{figure}

\begin{table*}[t!]
\centering
\resizebox{1.00\textwidth}{!}{
  \begin{tabular}{l|cc|cc|cc|cc|cc}
    \toprule
    \multirow{2}{*}{\textbf{Method}} & \multicolumn{2}{c|}{\textbf{Acc@1} ($\uparrow$)} & \multicolumn{2}{c|}{\textbf{Acc@3} ($\uparrow$)} & \multicolumn{2}{c|}{\textbf{Acc@5} ($\uparrow$)} & \multicolumn{2}{c|}{\textbf{\text{JSD} ($\downarrow$)}} & \multicolumn{2}{c}{\textbf{\text{HD} ($\downarrow$)}} \\
    \cmidrule{2-11} 
     & Location & Type & Location & Type & Location & Type & Location & Type & Location & Type \\
    \midrule
     Qwen2-Audio-7B (Text only) \cite{chu2024qwen2} & 6.68 & 12.00 & 16.39 & 53.56 & 18.68 & 76.96 & 0.2862 & 0.4346 & 0.5704 & 0.7150 \\
     Qwen2-Audio-7B (Text and Speech) \cite{chu2024qwen2} & \underline{7.36} & 12.83 & \underline{18.05} & \underline{57.60} & 18.41 & \underline{80.17} & 0.2353 & 0.4338 & 0.5188 & 0.7023 \\
     Qwen2.5-Omni-7B (Text only) \cite{xu2025qwen25omnitechnicalreport} & 6.53 & 12.47 & 12.11 & 43.94 & 27.32 & 69.90 & 0.2213 & 0.3334 & 0.4967 & 0.6138 \\
     Qwen2.5-Omni-7B (Text and Speech) \cite{xu2025qwen25omnitechnicalreport} & \underline{7.36} & \underline{20.07} & 12.23 & 56.41 & \underline{28.86} & 75.30 & \underline{0.2155} & \underline{0.3307} & \underline{0.4961} & \underline{0.6096} \\
    \midrule
    Affectron-330M (Proposed) & \textbf{12.59} & \textbf{75.77} & \textbf{29.69} & \textbf{85.99} & \textbf{44.06} & \textbf{91.69} & \textbf{0.0523} & \textbf{0.0051} & \textbf{0.2414} & \textbf{0.0723} \\
    \bottomrule
  \end{tabular}}
\caption{Comparison of zero-shot NV type and location prediction across LLMs and the Affectron model. Prediction accuracy is measured using top-$K$ accuracy (Acc@1/3/5) and distributional alignment is assessed using Jensen-Shannon Divergence (JSD) and Hellinger Distance (HD).}
\label{tab:mmllm_comparision}\vspace{-0.2cm}
\end{table*}
\section{LLM Baselines and Prompting Strategy}
\label{appendix:LLM}
Building on previous research regarding NV-aligned emotional corpora \cite{xin2024jvnv} and filler placement \cite{kim-etal-2025-fillerspeech}, we followed large language model (LLM) baselines and a prompting strategy. LLMs can partially capture discourse-level regularities associated with NV events, even in the absence of explicit acoustic supervision. For baseline comparisons with the proposed method, four widely adopted LLMs were selected to represent a range of sizes, architectures, and conversational capabilities:
\begin{itemize}
    \item \textbf{Qwen 2.5-7B}\footnote{\url{https://huggingface.co/Qwen/Qwen2.5-7B-Instruct}} \cite{hui2024qwen2}: an instruction-tuned multilingual model with robust instruction-following capabilities.
    \item \textbf{LLaMA 3.1-8B}\footnote{\url{https://huggingface.co/meta-llama/Llama-3.1-8B}} \cite{dubey2024llama}: a Transformer-based autoregressive language model designed for general-purpose text generation.
    \item \textbf{Vicuna-7B}\footnote{\url{https://huggingface.co/lmsys/vicuna-7b-v1.5}} \cite{chiang2023vicuna}: a dialogue-centric model fine-tuned using large-scale human conversational data.
    \item \textbf{GPT-oss-20B}\footnote{\url{https://huggingface.co/openai/gpt-oss-20b}} \cite{agarwal2025gpt}: a larger open-weight model with advanced long-context reasoning capabilities.
\end{itemize}

In all cases, a unified prompting scheme was employed for zero-shot NV type and location prediction, as illustrated in Figure \ref{fig:template}. For type prediction, each model received the transcript of a target utterance and a fixed list of NV categories, and was instructed to select the most appropriate NV type based solely on textual context. For location prediction, the utterance was presented as a sequence of word tokens indexed by their locations, and the model was prompted to select the most suitable insertion index for a single NV event. This experimental setup enabled a direct and controlled comparison between text-only LLM baselines and the proposed framework.

\section{Comparison with Multimodal Audio-LLM Baselines on NV Type and Location Prediction}
\label{appendix:multimodal_llm}
The comparison in Section~6.4 was originally designed to evaluate Affectron against prior LLM-based augmentation strategies, which are predominantly text-driven \cite{xin2024jvnv, kim-etal-2025-fillerspeech}. 
In that setup, text-only LLMs were prompted to predict NV type and insertion location from transcripts alone, whereas Affectron used speech-conditioned affective representations derived from the utterance and NV candidates.
To address modality fairness, we additionally evaluated multimodal audio-capable LLMs under both text-only and text-and-speech input settings.
For the multimodal fairness comparison, we included two open-source audio-capable LLMs:
\begin{itemize}
    \item \textbf{Qwen2-Audio-7B}\footnote{\url{https://huggingface.co/Qwen/Qwen2-Audio-7B-Instruct}} \cite{chu2024qwen2}: an instruction-tuned multimodal audio language model that supports both spoken audio and text inputs for speech understanding and generation tasks.
    \item \textbf{Qwen2.5-Omni-7B}\footnote{\url{https://huggingface.co/Qwen/Qwen2.5-Omni-7B}} \cite{xu2025qwen25omnitechnicalreport}: an open-source omni-modal language model designed to process text and audio inputs in a unified framework, with improved multimodal reasoning and speech interaction capabilities.
\end{itemize}

Table~\ref{tab:mmllm_comparision} summarize the results for NV location and NV type prediction, respectively. Incorporating audio inputs yielded only modest improvements over text-only settings for both multimodal LLMs. In contrast, Affectron consistently outperformed all multimodal baselines by a large margin across both top-$K$ accuracy and distributional alignment metrics. These results suggest that the gains of Affectron cannot be explained solely by access to acoustic cues, but instead arise from the explicit affect-aware modeling of NV selection and placement. Even when provided with audio input, general-purpose multimodal LLMs showed limited ability to capture fine-grained NV type and location distributions, highlighting the benefit of a dedicated affect-driven framework.

\section{Potential Risks}
Advances in speech synthesis have substantially expanded the capabilities of expressive and natural speech generation. At the same time, such progress introduces the possibility of misuse, particularly in scenarios where highly realistic synthesized speech could be leveraged to create deceptive or misleading audio content. These risks include the generation of fabricated speech for impersonation or misinformation, with potential societal consequences. Addressing such concerns requires complementary safeguards, including synthesized speech detection and watermarking, to support authentication, traceability, and responsible deployment.

\section{AI Assistance}
GPT-5.2 assisted with language polishing, including proofreading and grammatical corrections. The model was not involved in generating technical content, experimental design, or scientific claims.

\begin{figure*}[!t]
  \centering
\centerline{\includegraphics[width=0.99\textwidth]{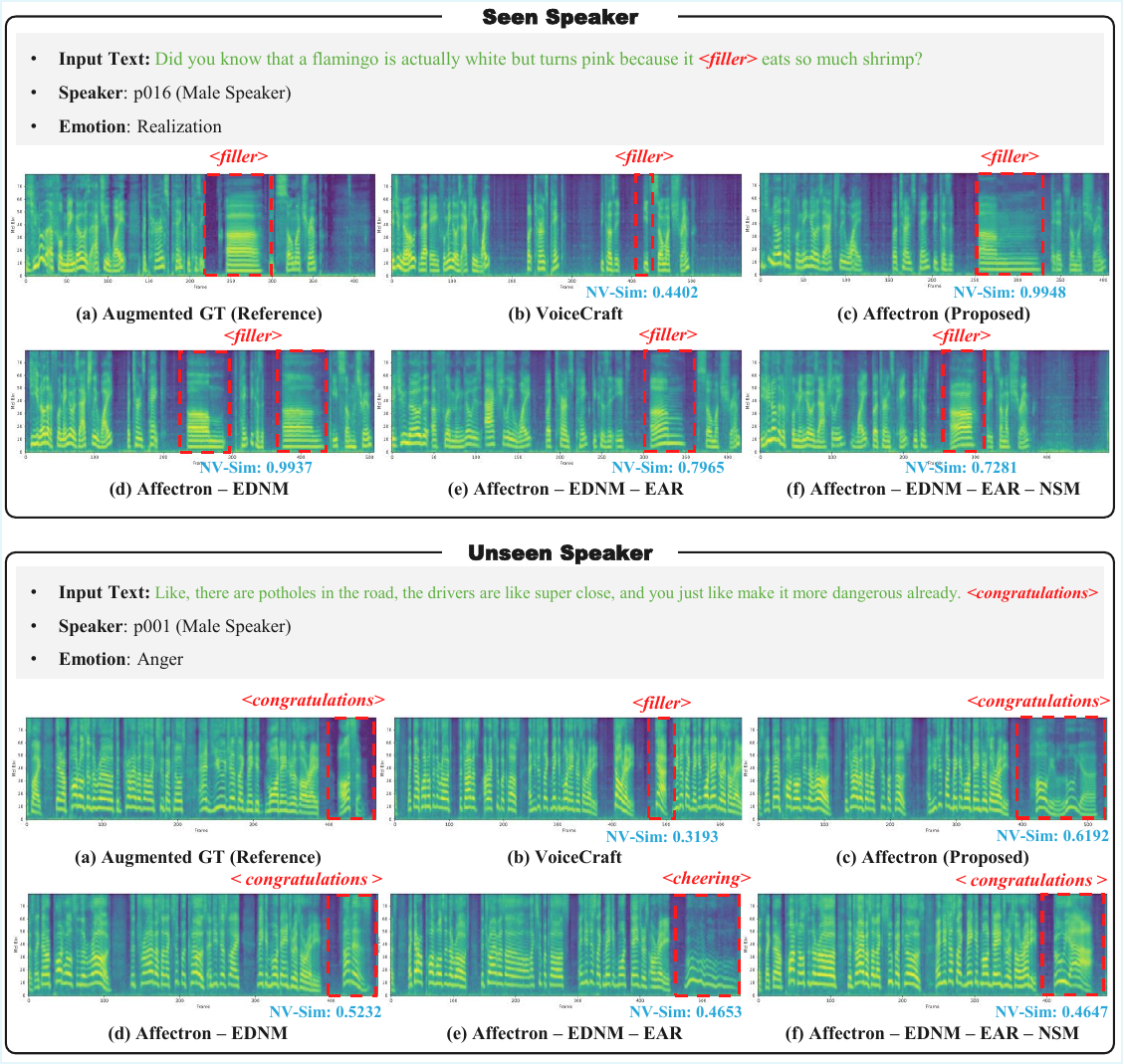}}\vspace{-0.2cm}
\caption{Comparison of Mel-spectrograms generated by different models for the same utterance. Results are shown for a seen speaker and an unseen speaker. NV segments are highlighted in red, and NV‑Sim scores are annotated in blue. EDNM, EAR, and NSM denote emotion-driven top-$K$ NV matching, emotion-aware top-$K$ routing, and NV structural masking, respectively. Augmented GT applies our affectively aligned NV augmentation to the ground truth.} \vspace{+0.2cm}
\label{melplot}
\end{figure*}

\section{Visualization Analysis of NV Expressive Generation}
Figure \ref{melplot} compares the Mel-spectrograms generated by Affectron, baseline models, and ablated variants under identical text and NV tag conditions. From this comparison, we made the following observations. First, Affectron consistently generated Mel-spectrograms that are well aligned with the given NV tags, accurately realizing NV events at the specified locations. In contrast, several baseline and ablated models either produced NVs that do not correspond to the provided tags or generated NVs at incorrect locations. Second, Affectron generated NV expressions that are emotionally aligned with adjacent verbal segments, resulting in smoother emotional transitions. This behavior was reflected in the higher NV-Sim scores achieved by Affectron. By comparison, most baseline and ablated models failed to maintain a consistent affective context between verbal speech and NVs, resulting in fragmented or incongruent expressive patterns. Overall, these visualizations demonstrated that Affectron generates NVs that follow the given tags and integrate them coherently with the surrounding verbal speech. For a broader set of qualitative examples and audio demonstrations, readers are referred to \url{https://choddeok.github.io/Affectron/}.

\end{document}